\documentclass[review]{elsarticle}

\usepackage{lineno,hyperref}
\usepackage{multirow}
\usepackage{longtable}
\usepackage{graphicx}

\usepackage{algorithm}
\usepackage{algpseudocode}
\usepackage{float}
\usepackage{tikz}
\usepackage{amssymb}

\usepackage{natbib} 
\usepackage{mathtools}
\usepackage[all,cmtip]{xy}

\usepackage{subcaption}

\journal{Computers \& Operations Research}

\biboptions{authoryear}

\newtheorem{definition}{\bf Definition}
\newtheorem{lemma}[definition]{\bf Lemma}
\newtheorem{theorem}[definition]{\bf Theorem}
\newtheorem{proposition}[definition]{\bf Proposition}
\newtheorem{corollary}[definition]{\bf Corollary}

\newenvironment{proof}{\noindent \textit{Proof. }}{\qed}

\long\def\comment#1\endcomment{}

\begin{document}

\begin{frontmatter}

\title{Algorithms for the Global Domination Problem} 

\author{Ernesto Parra Inza\fnref{myfootnote1}}
\ead{eparrainza@gmail.com}
\fntext[myfootnote1]{ Centro de Investigación en Ciencias, UAEMor, Cuernavaca, Morelos, México.}

\author[myfootnote1]{Nodari Vakhania\corref{mycorrespondingauthor}}
\cortext[mycorrespondingauthor]{Corresponding author}
\ead{nodari@uaem.mx}

\author{José María Sigarreta Almira\fnref{myfootnote2}}
\ead{josemariasigarretaalmira@hotmail.com}
\fntext[myfootnote2]{Facultad de Matemáticas, UAGro, Acapulco de Juárez, Guerrero, México.}

\author{Frank Ángel Hernández Mira\fnref{myfootnote3}}
\ead{fmira8906@gmail.com}
\fntext[myfootnote3]{Centro de Ciencias de Desarrollo Regional, UAGro, Acapulco de Juárez, Guerrero, México.}


\begin{abstract}
A dominating set $D$ in a graph $G$ is a subset of its vertices such that every vertex 
of the graph which does not belong to set $D$ is adjacent to at least one vertex from
set $D$. A set of vertices of graph $G$ is a global dominating set if it is a dominating set
for both, graph $G$ and its complement. The objective is to  find a global dominating set with
the minimum cardinality. The problem is known to be $NP$-hard. Neither exact nor approximation algorithm existed . We propose two
exact solution methods, one of them being based on an integer linear program (ILP) 
formulation, three heuristic algorithms and a special purification procedure 
that further reduces the size of a global dominated set delivered by any of
our heuristic algorithms. We show that the problem remains $NP$-hard for restricted 
types of graphs and specify some  families of graphs for which the heuristics 
guarantee the optimality. The second exact algorithm turned out to be
about twice faster than ILP for graphs with more than 230 vertices and up to 1080 vertices, 
which were the largest benchmark instances that were solved optimally. The heuristics 
were tested for the existing 2284 benchmark problem instances with up to 14000 vertices and 
delivered solutions for the largest instances  in less than one minute. Remarkably, for about 52\%  of the 1000 instances with the obtained optimal solutions, at least one of the heuristics 
generated an optimal solution, where the average approximation error for the remaining instances was 1.07\%. 
\end{abstract}


\end{frontmatter}


\section{Introduction}
A dominating set $D$ in a graph $G$ is a subset of its vertices such that every vertex
of the graph which does not belong to set $D$ is adjacent to at least one vertex from
set $D$. A set of vertices of graph $G$ is a global dominating set if it is a dominant set
for both, graph $G$ and its complement. The objective is to  find a global dominating set with
the minimum cardinality, an {\em optimal} global dominating set. The global domination 
problem was introduced by \cite{Sampathkumar} and was dealt with in a 
more general form by \cite{Dutton}, where it was shown that the problem is $NP$-hard.  
\cite{Haynes,Enciso} and \cite{Dutton} suggested some lower and upper bounds for the problem. 
To the best of our knowledge, there are known no other results on the problem. In 
particular, neither exact nor approximation algorithms have been earlier suggested. 

In this paper, 
we propose two exact solutions methods, one of them being based on an integer linear program (ILP) formulation, three heuristic algorithms and a special purification procedure that further reduces the size of a global dominated set delivered by any of our heuristic algorithms. We show that the 
problem remains $NP$-hard for planar and split graphs. We also give some families of graphs for 
which our heuristic algorithms obtain an optimal solution. These graph families containing 
star graphs as sub-graphs (see, for example,  \cite{social2012}) have immediate practical 
applications in social networks. In 
recent research, \cite{wang2009, wang2011} and \cite{abu2018} use networks to represent 
individuals with specific unfavorable social habits, such are drinking, smoking, drug consumption, alcoholism etc. A dominating set in such a network,  the so-called minimum 
positive influence dominating set (PIDS), represents a selected group of the most 
``influential'' individuals with the above social habits (ones have more 
followers in the social network).  Social intervention 
programs designed for the individuals with these habits are particularly directed to the
individuals from PIDS since due to budget constraints, it is not possible
to include all the individuals from the network. 
This model does not take into account a 
constant variation of the ``influence'' of some individuals, that occur in practice. A more 
flexible model, that takes into account a possible loss of the influence (followers) of the 
individuals from a given dominating set, would deal with a global dominating set instead of a 
dominating set.

We tested our 
algorithms for over 2200 known benchmark instances. Below we give a brief description 
of the experimental results (a detailed experimental analysis is presented in Section 9). 
We were able to obtain optimal solutions for the 1000 benchmark instances with up to 1086 
vertices from \cite{Parra2022}. In 
general, ILP formulation gave better execution  times for very dense graphs (ones with the
densities 0.8-0.9), whereas the other exact algorithm turned out to be about twice faster than 
the ILP method for moderate and large sized less dense graphs (ones with more than 230 vertices).

Using the optimal solutions obtained by our exact algorithms and the earlier known upper 
bounds for the problem, we evaluated the practical performance of the proposed three heuristics,
which were tested for the existing 2284 benchmark problem instances with up to 14000 vertices. 
We obtained solutions for the largest instances in less than one minute, where the heuristics 
halted within a few seconds for all instances with up to 1000 vertices. Remarkably, for about 52\%
	of the 1000 instances with the above obtained optimal solutions, at least one of the heuristics
	generated an optimal solution, where the average approximation error for the remaining instances
	was 1.07\%. In average over all the tested instances, the number of the 
vertices in the obtained solutions turned out to be less than 1/10th of the value of the
minimum upper bound.  These results were obtained after the application of the
proposed purification procedure, which turned out to be essential. For example, for 100\% of 
the analyzed instances, the solutions delivered by one of the heuristics were purified. 

The paper is organized as follows. In sections 2 and 3 we give basic definitions and we briefly 
overview the earlier known upper and lower bounds for the problem, and we study the 
computational complexity of the problem for general and some types of restricted graphs.
In Section 4 we describe an ILP formulation of the global domination problem and our 
exact algorithm. In Sections 5, 6 and 7 we describe the 
three heuristic algorithms and we establish graph families for which our 
heuristics give an optimal solution. The purification procedure is described 
in Section 8. Section 9 contains the description of
our experimental results, and Section 10 is the conclusion.

\section{Preliminaries}
 
We start this section with necessary notions and notations. 
Let $G=(V,E)$ be a simple graph of \emph{order} $|V|=n$ and \emph{size} $|E|=m$. 
For nonempty set $D\subseteq V$, and a vertex $v \in V$, $N_D(v)$ denotes the set of neighbors
of $v$ in $D$ and $N_D[v]=N_D(v)\cup\{v\}$. The \emph{degree} of $v$ in $D$ will be denoted by
$\delta_D(v) = |N_D(v)|$. Analogously, $\overline{N}_D(v)$ denotes the set of non  neighbors
of $v$ in $D$, $\overline{N}_D[v]=\overline{N}_D(v)\cup\{v\}$ and $\overline{\delta}_D(v)$ denotes
the cardinality of the set $\overline{N}_D(v)$ (i.e $\overline{\delta}_D(v) = |\overline{N}_D(v)|$) .
For short, we will often use $N(v)$, $N[v]$, $\delta(v)$,  $\overline{N}(v)$ and $\overline{N}[v]$
instead of $N_V(v)$, $N_V[v]$, $\delta_V(v)$, $\overline{N}_V(v)$ and $\overline{N}_V[v]$,
respectively. Therefore, $N_A(B) = \bigcup_{v\in B}N_A(v)$ and $N_A[B] = \bigcup_{v\in B}N_A[v]$,
where $A$ and $B$ are subsets of $V$, for short if $A=V$, then $N(B) = \bigcup_{v\in B}N(v)$ and
$N[B] = \bigcup_{v\in B}N[v]$. The \emph{private neighborhood} $pn(v,S)$ of vertex $v \in S\subseteq V$
is defined by $pn(v,S)=\{u \in V: N(u)\cap S = \{v\}\}$. A vertex in $pn(v,S)$ is called a
\emph{private neighbor} of $v$.

A nonempty subset $D \subseteq V$ is called a global dominating set in $G$ (or GDS 
set for short) if $D$ is a dominating set of both $G$ and $\overline{G}$, where 
$\overline{G}$ is the complement of the graph G, i.e., the graph consisting of
all vertices of $G$ and all the edges which are not in set $E=E(G)$; in other 
words, for every vertex  $v\in V\setminus D$, (i) there is at least one vertex $x\in D$
such that $(x,v)\in E$, and (ii) there is at least one vertex $y\in D$ such that 
$(y,v)\not\in E$. The {\em domination number} of graph $G$,  $\gamma(G)$ is the
number of vertices in a dominating set of minimum cardinality of this graph. 
We will refer to a dominating set of graph $G$ with cardinality $\gamma(G)$ as 
$\gamma(G)$-set. The {\em global domination number} $\gamma_g(G)$ 
and $\gamma_g(G)$-set are defined similarly for the global domination problem.

If $G_1$, $G_2$, \ldots , $G_r$ with $r\ge 2$, are the connected components of a 
graph $G$, then any minimum GDS in $G$ is formed by a minimum dominating set in 
each subgraph. This implies that, $\gamma_g(G)= \sum_{i=1}^r\gamma(G_i)$. 
Furthermore, it is known that $\gamma_g(G)= \gamma_g(\overline{G})$. Due to these
observations, we assume that our graphs and their complements are connected.

The \emph{diameter} $d(G)$ of graph $G$ is the maximum number of edges on a shortest path 
between a pair of vertices, and the \emph{radius} $r(G)$ is the minimum number of edges on 
a shortest path between a pair of vertices in that graph. A {\em leaf} vertex is a degree one 
vertex, and a {\it support} vertex is a vertex adjacent to a leaf.  The set of support vertices
will be denoted by  $Supp(G)$, and the degree of a vertex with the maximum number of neighbors 
in graph $G$ by $\Delta(G)$.

\smallskip

Now we briefly overview the known lower and upper limits for the problem.  
We obtain our overall lower bound $L$ using the following known results.
\begin{theorem}[\cite{haynes2017domination}]\label{teorema1}
	If $G(V,E)$ is a connected graph of order $n$, then
	\begin{itemize}
		\item $ \gamma(G)\geq\frac{n}{\Delta(G)+1}$,		
		\item $ \gamma(G)\geq \frac{2r(G)}{3}$,		
		\item $ \gamma(G)\geq \frac{d(G)+1}{3}$, and		
		\item $ \gamma(G) \geq |Supp(G)|$.
	\end{itemize} 
\end{theorem}

\begin{theorem}[\cite{Haynes}]\label{teorema4}
	If $G(V,E)$ is a connected graph, $ \gamma_g(G)\geq \gamma(G)$.
\end{theorem}

The next result is an immediate consequence of the Theorems \ref{teorema1} and \ref{teorema4}:

\begin{corollary}\label{cororario1}
	Let $G(V,E)$ be a connected graph of order $n$. Then $$L=\max\left\{\frac{n}{\Delta(G)+1},\frac{2r(G)}{3},\frac{d(G)+1}{3},|Supp(G)|\right\}$$ 
	is a lower bound of $\gamma_g(G)$.
\end{corollary}

The overall upper bound $U$ is obtained using  the following known results.

\begin{theorem}[\cite{Dutton}]\label{teorema5}
	If $G(V,E)$ is a connected graph of maximum degree $\Delta(G)$, then $ \gamma_g(G)\leq U_1 = \min\{\Delta(G),\Delta(\overline{G})\} + 1$.
\end{theorem}

\begin{theorem}[\cite{Dutton}]\label{teorema6}
	For any graph $G(V,E)$ of minimum degree $\delta(G)$, $ \gamma_g(G)\leq U_2 = \delta(G)+2$ if 
	$\delta(G)=\delta(\overline{G})\leq 2$; otherwise $ \gamma_g(G)\leq U_2 = \max\{\delta(G),\delta(\overline{G})\}+1$.
\end{theorem} 

We summarize Theorems \ref{teorema5} and \ref{teorema6} in the next corollary:

\begin{corollary}\label{cororario2}
	Let $G(V,E)$ be a connected graph. Then $U=\min\{U_1,U_2\}$ is an upper bound of $\gamma_g(G)$.
\end{corollary}

\section{NP-hardness results}

As already noted, the global domination concept was intruded by \cite{Sampathkumar}. 
\cite{Dutton}  introduced the so-called factor domination problem, and showed 
that the global domination problem can be restricted to the domination problem,
which is known to be NP-hard  (\cite{Garey}). We give a few definitions from 
\cite{Dutton} and then we give a scratch of this intuitive reduction.

Let us say that graph $G=(V, E)$ can be decomposed into $k-factors$ $G_1,\dots,G_k$, $G_i =(V_i, E_i)$ 
if for each graph $G_i$, $V_i = V$ and the collection of all edges $V_1,\dots,V_k$ forms a partition
of set $E$. 
Given a decomposition of graph $G=(V, E)$ into the $k$-factors $G_1,\dots,G_k$,  $D \subseteq V$
is called a {\em factor dominating set} if $D$ is a dominating set for each $G_i$, $i=1,\dots,k$. The 
{\em factor domination number}  $\gamma(G_1,\dots,G_k)$  is the number of elements in a smallest factor dominating set.

Let us now consider a 2-factoring of a complete graph $K_n$, and any its sub-graph $G=(V,E)$ 
with $|V|=n$. Let $G_1=G$ and $G_2=\bar G$.  Then note that $G_1,G_2$ is a 2-factorization 
of graph $K_n$. Then we immediately obtain that the factor domination problem for graph $K_n$
and with $k=2$ is equivalent to the global domination problem in that graph. Note now that
factor domination problem with $k=1$ is the standard domination problem, which is NP-hard. 
Hence,  the factor domination problem is also NP-hard also for any $k\ge 2$, in particular, 
for $k=2$:

\begin{theorem} [\cite{Dutton}]
	The global domination problem is NP-hard.
\end{theorem}

In the remaining part of this section, we show that the domination problem in restricted types of 
graphs remains NP-hard. In particular, we consider split and planar graphs (\cite{merris2003}, \cite{MEEK}).  
We will use $D$ ($D_g$) for a minimum (global) dominating set.

\subsection{Split graphs}

A {\it split graph} is a graph in which the vertices can be partitioned into a clique 
$C$ and an independent set $I$ (see \cite{Foldes}, \cite{Tyshkevich}, \cite{COLOMBI}, \cite{JOVANOVIC}). Let $G$ be a 
connected split graph and let $C$ ($I$, respectively) be a maximum clique (the independent 
set, respectively) in that graph (recall that  our graphs are assumed to be connected 
due to our earlier remark). Since graph $G$ is connected, there exists an edge from any vertex 
from set $I$ to a vertex from set $C$. In particular, the following proposition holds. 

\begin{proposition}
	If $G$ is a connected split graph, its complement $\bar G$ is also a connected split graph.
\end{proposition}
\begin{proof}
	Note that the roles of sets $C$ and $I$ are interchanged in the complement $\bar G$, hence $\bar G$ is also a split graph. The complement is also connected, since in graph $G$, for any vertex from set $I$, there exists at least one vertex from set $C$ which is not adjacent to that vertex. 
\end{proof}

\begin{lemma}[\cite{Bertossi}]\label{Be}
	Suppose $G$ is a connected split graph with maximum clique $C$ and independent set $I$.
	Then any minimum dominating set $D$ of graph $G$ is a (not necessarily proper) subset
	of set $C$. 
\end{lemma}

\begin{lemma}\label{cor} 
	Suppose $G$ is a connected split graph with maximal clique $C$, independent set $I$
	and a minimum dominating set $D$. Then if $D=C$ then $D_g=D$.
\end{lemma}

\begin{proof}
	Since $D=C$, for every vertex  $y\in I$, there must exist a vertex $z\in C$ which is not
	associated with vertex $y$. Then vertex $z$ will dominate vertex $y$ in the complement 
	$\bar G$. Hence, $D$ is also a global dominating set. 
\end{proof}

\begin{theorem}
	Given a connected split graph $G$, any global dominating set for this graph is formed
	either by the dominating set of that graph or by the dominating set of that 
	graph and one (or more)  vertices  from set $I$. Therefore, the global domination 
	problem in graph $G$ reduces to the domination problem in that graph, and hence it is NP-hard. 
\end{theorem}

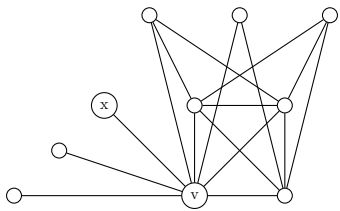
\begin{figure}[h!]
	\centering
	\begin{tikzpicture}[scale=.6, transform shape]

		\node [draw, shape=circle] (s1) at  (0,0) {v};
		\node [draw, shape=circle] (s2) at  (0,2) {};
		\node [draw, shape=circle] (s3) at  (2,2) {};
		\node [draw, shape=circle] (s4) at  (2,0) {};
		
		\node [draw, shape=circle] (x) at  (-2,2) {x};
		\node [draw, shape=circle] (x1) at  (-3,1) {};
		\node [draw, shape=circle] (x2) at  (-4,0) {};
		\node [draw, shape=circle] (s5) at  (-1,4) {};
		\node [draw, shape=circle] (s6) at  (3,4) {};
		\node [draw, shape=circle] (s7) at  (1,4) {};
		
		\draw(s1)--(x); \draw(s1)--(x1); \draw(s1)--(x2);
		\draw(s1)--(s2)--(s3)--(s4)--(s1)--(s3);
		\draw(s4)--(s2);
		
		\draw(s1)--(s7)--(s4);
		\draw(s1)--(s5)--(s2); \draw(s3)--(s5);
		\draw(s2)--(s6)--(s4); \draw(s3)--(s6);
		

	\end{tikzpicture}
	\caption{\small A split graph with both, the type $x$ and the type $u$ jobs}\label{fig-split}
\end{figure}

\begin{proof}
	Let us consider two basic types of a split graph $G$. In the first type of split graph 
	$G$, every vertex $v\in C$ is associated with at least one grade one (pendant) vertex 
	$x\in I$ (i.e., $x$ is a private neighbor of $v$). For such a $G$, there is 
	only one minimum dominating set $D=C$. Set $D$ is also a global dominating set for graph 
	$G$ by Lemma \ref{cor}, and the first claim in the theorem holds. Suppose now $G$ is of 
	the second type, i.e., there is a vertex in set $I$ associated with  two or more vertices 
	in set $C$, see Figure \ref{fig-split} (note that, since graph $G$ 
	is connected, every vertex in set $I$ is associated with at least one vertex of set $C$).
	Let $u\in C$ be a vertex with no private neighbor in set $I$. (In the graph of 
	Figure \ref{fig-split}, the clique $C$ is formed by the four vertices of the rectangular
	from the lower right part of that figure. This clique contains one vertex of type $v$
	and three vertices of type $u$.) It is easy to see that a minimum dominating set $D$ of 
	graph $G$ is a proper subset of set $C$ (see Lemma \ref{Be}). Indeed, since there is a 
	vertex in set $I$ associated with two or more (type $u$) vertices in set $C$, set $D$ 
	is formed by the vertices of clique $C$ but one (or more) type $u$ vertices. Note that
	these type $u$ vertices will not be dominated by set $D$ in the complement $\bar G$. 
	Hence, to form a global dominating set, we need to add to set $D$ one (or more) vertices 
	from set $I$, those, which are not adjacent to the above type $u$ vertices from set 
	$C\setminus D$.
	
	We showed that the global dominating set problem 
	in split graphs reduces to the standard domination problem for both types of the split 
	graphs. The latter problem is known to be NP-hard  \cite{Bertossi}. The theorem is proved. 
\end{proof}

\subsection{Planar graphs}

In \cite{Enciso} the global domination problem in planar graphs was studied. In 
particular, upper bounds for special types of planar graphs were obtained. The
complexity status of the problem, however,  has not been yet addressed, neither in 
the above reference nor in any other reference that we know. Here we prove that 
the global domination problem in planar graphs is NP-hard.

\begin{figure}[H]
	\centering
	\begin{tikzpicture}[scale=.7, transform shape]
		
		\node [draw, shape=circle] (t1) at  (-4,2) {};
		\node [draw, shape=circle] (t2) at  (-4.5,3) {};
		\node [draw, shape=circle] (t3) at  (-3,3) {};
		
		\node [draw, shape=circle] (st2) at  (-5,2) {};
		\node [draw, shape=circle] (st3) at  (-3,2) {};
		
		\draw(t2)--(t1)--(t3);
		\draw(t2)--(st2)--(t1)--(st3)--(t3);
		
		\node [left] at (-1.5,1.5) {\large ($ G $)};
		\node [left] at (4.5,1.5) {\large ($ H_i $)};

		\node [draw, shape=circle] (s1) at  (2,1.8) {};
		\node [draw, shape=circle] (s2) at  (0.5,2) {};
		\node [draw, shape=circle] (s3) at  (1.5,3) {$ w$};
		\node [draw, shape=circle] (s4) at  (1,4.2) {};
		\node [draw, shape=circle] (s5) at  (2.5,4) {};
		\node [draw, shape=circle] (s6) at  (0.2,3) {};
		\node [draw, shape=circle] (s7) at  (2.9,2.8) {$ r $};
		
		\draw(s3)--(s1);
		\draw(s3)--(s2);
		\draw(s3)--(s4);
		\draw(s3)--(s5);
		\draw(s3)--(s6);
		\draw(s3)--(s7);

	\end{tikzpicture}
	\caption{A star graph $H$ and a planar graph $G$}
	\label{G-H}
\end{figure}
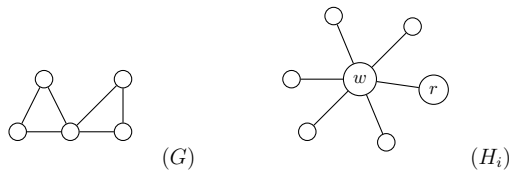

\begin{figure}[H]
	\centering
	\begin{tikzpicture}[scale=.6, transform shape]
		
		\node [draw, shape=circle] (s11) at  (2,1.8) {};
		\node [draw, shape=circle] (s21) at  (0.5,2) {};
		\node [draw, shape=circle] (s31) at  (1.5,3) {$ w_1 $};
		\node [draw, shape=circle] (s41) at  (1,4.2) {};
		\node [draw, shape=circle] (s51) at  (2.5,4) {};
		\node [draw, shape=circle] (s61) at  (0.2,3) {};
		\node [draw, shape=circle] (s71) at  (2.9,2.8) {$ r_1 $};
		
		\draw(s31)--(s11);
		\draw(s31)--(s21);
		\draw(s31)--(s41);
		\draw(s31)--(s51);
		\draw(s31)--(s61);
		\draw(s31)--(s71);
		
		\node [draw, shape=circle] (s12) at  (2,-1.3) {};
		\node [draw, shape=circle] (s22) at  (0.5,-1) {};
		\node [draw, shape=circle] (s32) at  (1.5,0) {$ w_2 $};
		\node [draw, shape=circle] (s42) at  (1,1.2) {};
		\node [draw, shape=circle] (s52) at  (2.5,1) {};
		\node [draw, shape=circle] (s62) at  (0.2,0) {};
		\node [draw, shape=circle] (s72) at  (2.9,-0.3) {$ r_2 $};
		
		\draw(s32)--(s12);
		\draw(s32)--(s22);
		\draw(s32)--(s42);
		\draw(s32)--(s52);
		\draw(s32)--(s62);
		\draw(s32)--(s72);

		\node [draw, shape=circle] (1) at  (4.9,2) {$ _1 $};
		\node [draw, shape=circle] (2) at  (4.9,0.5) {$ _2 $};
		\node [draw, shape=circle] (3) at  (6.5,1.5) {$ _3 $};
		\node [draw, shape=circle] (4) at  (8.1,0.5) {$ _4 $};
		\node [draw, shape=circle] (5) at  (8.1,2) {$ _5 $};
		
		\draw(5)--(3)--(1);
		\draw(5)--(4)--(3)--(2)--(1);
		
		\draw(s71)--(1);
		\draw(s72)--(2);
		
		\node [draw, shape=circle] (s15) at  (12,1.8) {};
		\node [draw, shape=circle] (s25) at  (10.5,2) {};
		\node [draw, shape=circle] (s35) at  (11.5,3) {$ w_5 $};
		\node [draw, shape=circle] (s45) at  (11,4.2) {};
		\node [draw, shape=circle] (s55) at  (12.5,4) {};
		\node [draw, shape=circle] (s65) at  (10,3) {$ r_5 $};
		\node [draw, shape=circle] (s75) at  (12.9,2.8) {};
		
		\draw(s35)--(s15);
		\draw(s35)--(s25);
		\draw(s35)--(s45);
		\draw(s35)--(s55);
		\draw(s35)--(s65);
		\draw(s35)--(s75);
		\draw(5)--(s65);
		
		\node [draw, shape=circle] (s14) at  (12,-1.3) {};
		\node [draw, shape=circle] (s24) at  (10.5,-1) {};
		\node [draw, shape=circle] (s34) at  (11.5,0) {$ w_4 $};
		\node [draw, shape=circle] (s44) at  (11,1.2) {};
		\node [draw, shape=circle] (s54) at  (12.5,1) {};
		\node [draw, shape=circle] (s64) at  (10,0) {$ r_4 $};
		\node [draw, shape=circle] (s74) at  (12.9,-0.3) {};
		
		\draw(s34)--(s14);
		\draw(s34)--(s24);
		\draw(s34)--(s44);
		\draw(s34)--(s54);
		\draw(s34)--(s64);
		\draw(s34)--(s74);
		\draw(4)--(s64);
		
		\node [draw, shape=circle] (s13) at  (7,3) {$ r_3 $};
		\node [draw, shape=circle] (s23) at  (5.5,3.2) {};
		\node [draw, shape=circle] (s33) at  (6.5,4.2) {$ w_3 $};
		\node [draw, shape=circle] (s43) at  (6,5.4) {};
		\node [draw, shape=circle] (s53) at  (7.5,5.2) {};
		\node [draw, shape=circle] (s63) at  (5,4.2) {};
		\node [draw, shape=circle] (s73) at  (7.9,4) {};
		
		\draw(s33)--(s13);
		\draw(s33)--(s23);
		\draw(s33)--(s43);
		\draw(s33)--(s53);
		\draw(s33)--(s63);
		\draw(s33)--(s73);
		\draw(3)--(s13);
		
	\end{tikzpicture}
	\caption{The rooted product results in a planar graph  $ G\circ_r H$}
	\label{G-H2}
\end{figure}
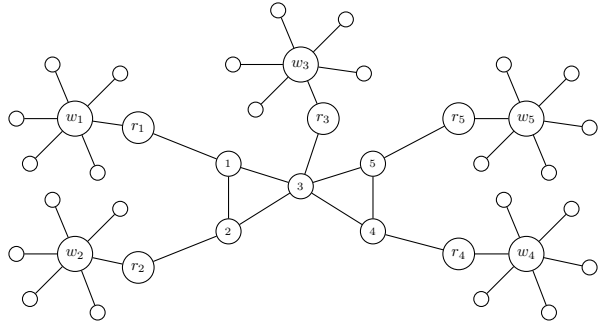

To start with, let us define the {\em rooted product} of two graphs. Given a graph $G$
with $n$ vertices and a graph (tree) $H$ with the root vertex $r$, the rooted product 
$G\circ_r H$ is the graph obtained from graphs $G$ and $H$ as follows. We extend every vertex
$v$ of graph $G$ with a copy of tree $H$, i.e., we associate with vertex $v$ vertex
$r$ extending in this way graph $G$. In  Figure \ref{G-H} we illustrate a  
planar graph $G$ and tree $H$. Figure \ref{G-H2} illustrates the rooted product of
these two graphs.

\begin{theorem}
	Given a planar graph $G$ of order $n$ and a star graph $H$ of order greater than two
	with support $w$, graph 
	$ G\circ_r H$ is planar. Moreover, the global dominating set of graph $ G\circ_r H$ 
	is the domination set of graph $G$ augmented with $n$  support vertices of the $n$ 
	copies of graph $H$. Therefore, the global domination problem in graph $G$ is $NP$-hard.
\end{theorem}

\begin{proof}
	Since both graphs $G$ and $H$ are planar, the rooted product of these graphs clearly 
	produces another planar graph. We prove the remaining part of the theorem. Let $D$ be
	a minimum dominating set of graph $G$. It is easy to see that 
	$D \cup \{w_1, w_2, \dots, w_n\}$ is a global dominating set of graph $ G\circ_r H$.
	Indeed, every vertex of $i$th copy $H_i$ of graph $H$ is dominated by  vertex $w_i$, 
	hence $D \cup \{w_1, w_2, \dots, w_n\}$ is a dominating set of graph $ G\circ_r H$. 
	$D \cup \{w_1, w_2, \dots, w_n\}$ is also a dominating set of the complement of 
	graph $ G\circ_r H$ since vertex $w_i$ will dominate any vertex from the complement 
	which is not in $H_i$, whereas all vertices of graph $H_i$ will be dominated by any 
	$w_j$, $j\ne i$. (Figure \ref{G-H2} represents the rooted product $ G\circ_r H$ of the
	two graphs of Figure \ref{G-H}, where set $D$ consists of vertex $3$.) 
	
	Note now  that 
	$D \cup \{w_1, w_2, \dots, w_n\}$ is a a minimum dominating set for graph $ G\circ_r H$
	since clearly, at least one vertex from each graph $H_i$ is to be included in any 
	dominating set of that graph. Hence,
	$D \cup \{w_1, w_2, \dots, w_n\}$ is the minimum global dominating set for graph 
	$ G\circ_r H$ since it coincides with the minimum dominating set of that graph. 
	
	The last claim in the theorem follows since the standard dominating set problem 
	in planar graphs is known to be $NP$-hard (\cite{Enciso}).
\end{proof}

\section{ Exact algorithms}
 
\subsection{ILP formulation of the problem}

Given graph $G=(V,E)$ of order $n$, its neighborhood matrix $A=(a_{ij})_{n\times n}$
is defined as follows: $ a_{ij} = a_{ji} = 1 $ if $(i,j) \in E(G)  $ or $ i=j $, and $ a_{ij} = a_{ji} = 0 $ otherwise. 
The neighborhood matrix of $\overline{G}$, 
$B=(b_{ij})_{n\times n}$, is defined as follows: $ b_{ij} = b_{ji} = 1 $ if $a_{ij} = 0 $ or $ i=j $, and $ b_{ij} = b_{ji} = 0 $ if $a_{ij} = 1 $ and $ i\neq j $.

Given a global dominating set $D$ of cardinality $\gamma_g(G)$ and $i \in V(G)$, we define our decision variables as follows:
\begin{equation}
	x_{i}= \left\{  \begin{array}{lcl}
		1, &  \mbox{ if }  x_{i} \in D\\
		0, &  \mbox{ otherwise, }   
	\end{array}
	\right.
\end{equation}
The global dominating set problem can now be formulated as follows:
\begin{equation} \label{fo}
	\min \sum_{i=1}^{n}x_{i}=\gamma_g(G),
\end{equation}
\begin{equation}\label{st1}
	subject \; to:\; \sum_{j=1}^{n}a_{ij}x_{j} \geq 1, \; x_{j}\in \lbrace 0,1\rbrace , \; \forall i \in V(G),
\end{equation}
\begin{equation}\label{st2}
 \;\;\;\;\;\;\;\;\;\;\;\;\;\;\;\;\;\;\;\;\sum_{j=1}^{n}b_{ij}x_{j} \geq 1, \; x_{j}\in \lbrace 0,1\rbrace , \; \forall i \in V(\overline G),
\end{equation}
Here the objective function is $ \sum_{i=1}^{n}x_{i}$, that we
wish to minimize (\ref{fo}). We have $2n$ restrictions (\ref{st1} and \ref{st2}), guarantying that  
for each $ j \in V (G) $ and each $ j \in V (\overline{G}) $, at least one of the vertices in $ D $ is adjacent to $ j $ or $ j \in D$.

\subsection{The implicit enumeration algorithm}

From here on, we shall refer to any global dominating set as a \textit{feasible 
solution}, whereas any subset of vertices from set $V$ will be referred to as a 
solution. A global dominating set with the minimum cardinality is {\em optimal} 
feasible solution. The implicit enumeration algorithm starts with a feasible
solution obtained by one of our heuristics H1, H2, and H3,
which  also defines the initial upper bound $U$ on the size of a feasible 
solution. Given the lower bound $L$ from Corollary \ref{cororario1} (see Section 8) 
and the upper bound $U$, we restrict our search for feasible solutions with the 
size in the range $[L,U]$ using binary search, similarly to the approach from 
\cite{ejor2023exact} used for the dominating set problem. 

The solutions of the size $\nu\in [L,U]$ are generated and tested for the feasibility 
based on the specially formed priority list of solutions. We use procedures $Procedure\_Priority\_LIST()$ and $Procedure\_Next()$ from \cite{ejor2023exact}
to create the priority lists and generate the solutions of size $\nu$. The sizes of feasible solutions are derived by binary division search accomplished in the interval 
$[L, U]$. For each created solution $\sigma$ of size $\nu$ for $Procedure\_Next()$, feasibility condition is verified, i.e., it is verified if the solution forms a global dominating set. 

Now we describe the general framework of the algorithm (which is based on the same
principles as the one from \cite{ejor2023exact}).
Let $\sigma$ be the current solution of size $\nu$ (initially, $\sigma$ is the best
feasible solution delivered by either of the Algorithms H1, H2 and H3: 

\begin{itemize}	
	\item If solution $\sigma$ is a global dominating set, then the current upper bound $U$ is updated to $\nu$.
	The algorithm proceeds with the next (smaller) trial value $\nu$   from the interval $[L,\nu)$ derived by binary search procedure. If all trial $\nu$s were already tested, then $\sigma$ is an optimal solution. The algorithm returns solution $\sigma$ and halts.
	
	\item If the current solution $\sigma$ of size $\nu$ is not a global dominating set, then  $Procedure\_Next(\nu)$ is called and the next to $\sigma$ solution of size $\nu$ from the corresponding priority list is tested.
		
	\item If $Procedure\_Next(\nu)$ returns $NIL$, {\it i.e.,}, all the solutions of size $\nu$ were already tested for the feasibility (none of them being feasible), the current lower bound $L$ is updated to $\nu$. The algorithm proceeds with the next (larger) trial value $\nu$  from the interval $[\nu, U)$ derived by  binary search procedure. If all trial $\nu$s were already tested, then $\sigma$ is an optimal solution. The algorithm returns solution $\sigma$ and halts. 
\end{itemize}

We conclude this section with a formal description of the algorithm. 
\medskip
\begin{algorithm}[H]
	\fontsize{7pt}{.25cm}\selectfont
	\caption{Algorithm\_BGDS}\label{alg1}
	\begin{algorithmic}
		\State \textbf{Input:} A graph $G$.
		\State \textbf{Output:} A $\gamma_g(G)$-$set$ $D$.
		
		\State $L := \max\{\frac{n}{\Delta(G)+1},\frac{2r}{3},\frac{d+1}{3},|Supp(G)|\}$;
		\State $D:=\sigma$; \hspace{.3cm} \{the best feasible solution delivered by the algorithms \ref{alg_one}, \ref{alg_two}, and \ref{alg_3}\}
		\State $U := |D|$;
		\State $\nu := \lfloor(L+3U)/4\rfloor$;
		
		\State $LIST := Procedure\_Priority\_LIST(G,D)$;
		
		\{ iterative step \}

		\While{$U-L>1$} 
		
		\If{$Procedure\_Next(\nu, LIST, G)$ returns $NIL$}
		\State $L := \nu$;
		\State $\nu := \lfloor(L+3U)/4\rfloor$;
		\Else  \hspace{.3cm} \{ A feasible solution was found ($\sigma_h(\nu)$)\}
		\State $D := \sigma_h(\nu)$;
		\State $U := \nu$;
		\State $\nu := \lfloor(L+3U)/4\rfloor$;
		\State $LIST := Procedure\_Priority\_LIST(G,D)$; 
		\EndIf
		
		\EndWhile
		\State \textbf{return} $D$;
	\end{algorithmic}
\end{algorithm} 
\bigskip

\section{Algorithm H1}

Our first heuristic Algorithm H1 constructs the destiny global dominating set
$D$ in a number of iterations; we denote by $D_h$ the set of vertices already 
constructed by iteration $h$. Given set $D_h$, let $A_h$ be the set of all
non-dominated vertices by iteration $h$, i.e., the set of vertices from 
$V\setminus D_h$ without a neighbor in $D_h$, $N(A_h)\cap D_h = \emptyset$. 
Let $B_h$ be the set of non-dominated vertices in $\overline G$, i.e., 
the set consisting of all vertices  $v\in V\setminus D_h$ such that 
there is an edge $(u,v)\in E$, for all $u\in D_h$. Note that sets $D_h$, $A_h$ 
and $B_h$ are disjoint. Let $C_h = V \setminus \{D_h\cup A_h\cup B_h\}$. It
follows that $C_h$ is the {\em globally dominated} set in iteration $h$, i.e., the
set of vertices $x\in V\setminus D_h$ such that there exist $y\in D_h$ with 
$(x,y)\in E$ and $z\in D_h$ with $(x,z)\not\in E$. The next lemma follows.

\begin{lemma}\label{lema1}
	The sets of vertices $A_h$, $B_h$, $C_h$ and $D_h$ are disjoint and  $A_h \cup B_h \cup C_h \cup D_h = V$.
\end{lemma}

The \emph{global active degree} in iteration $h\geq 1$ of vertex $v\in V\setminus D_{h-1}$,
$GAD(v)$, is the number of neighbors of $v$ in $A_{h-1}$ plus the number of non-neighbors 
of $v$ in $B_{h-1}$, i.e., 
$$GAD(v) = |N_{A_{h-1}}[v] \cup \overline{N}_{B_{h-1}}[v]|.$$ 

Algorithm H1 initiates at iteration 0 by determining two vertices $u$ and $v$ 
with 
$$|(N[u]\cup N[v])\setminus (N[u]\cap N[v])| =
 \max_{x,y\in V} |(N[x]\cup N[y])\setminus (N[x]\cap N[y])|$$

(note that $\gamma_g(G) \geq 2$.) In other words, we start with a pair of vertices 
globally dominating the maximum possible number of vertices. So, we let $D_0 = \{u,v\}$. Hence, $C_0 = (N[u]\cup N[v])\setminus (N[u]\cap N[v])$, $A_0 = V\setminus (N[u]\cup N[v])$ and $B_0 = V\setminus (D_0\cup C_0\cup A_0)$.  

\begin{lemma}\label{obsD0}
If $A_0=B_0=\emptyset$ then 

\begin{equation}\label{union}
(N[u]\cup N[v])\setminus (N[u]\cap N[v]) = V\setminus \{u,v\}, 
\end{equation}

and $D_0=\{u,v\}$ is an optimal feasible solution. 
\end{lemma} 
\begin{proof}
Equation (\ref{union}) immediately follows from the condition 
$A_0=B_0=\emptyset$ and the fact that  $|D_0|=2$. By equation (\ref{union}),  
the set of vertices $V\setminus \{u,v\}$  is partitioned into two (disjoint) 
sets $N[u]$ and $N[v]$. This implies that $D_0$ is a global dominating set. 
Alternatively, $A_0=B_0=\emptyset$ yields $C_0=V\setminus D_0$ 
by Lemma \ref{lema1}. Then $D_0$ is minimum since a global dominating set
cannot contain less than 2 vertices. 
\end{proof}

If the condition in Lemma \ref{obsD0} is not satisfied, then the algorithm
proceeds with iteration 1. It halts at iteration $h\ge 1$  
if $A_{h-1}=B_{h-1}=\emptyset$. Otherwise, it determines vertex 
$v_h \in  V\setminus D_{h-1}$ with the maximum global active degree and lets
$D_h:= D_{h-1} \cup \{v_h\}$. The sets $A_h$, $B_h$ and $C_h$ are updated 
correspondingly, see below  the pseudo-code for the details. 

\medskip

\begin{algorithm*}
	\caption{H1}\label{alg_one}
	\fontsize{7pt}{.25cm}\selectfont
	\begin{algorithmic}
		\State Input: A graph $G(V,E)$.
		\State Output: A global dominating set $D_h$.
		\State $h := 0$; 
		\State $\{u,v\}$ := two vertex with the largest possible cardinality of the set \\ $(N[u]\cup N[v])\setminus (N[u]\cap N[v])$;
		\State $D_0:= \{u,v\}$;  
		\State $C_0:= (N[u]\cup N[v])\setminus (N[u]\cap N[v])$;  
		\State $A_0:= V\setminus (N[u]\cup N[v])$; 
		\State $B_0:= V\setminus (D_0\cup C_0 \cup A_0)$;  
		
		\{ iterative step \} 
		
		\While{ $A_h \neq \emptyset \vee B_h \neq \emptyset$ } 
		\State $h := h+1$; 
		\State $v_h$ := a vertex in the set $V\setminus D_{h-1}$ with the maximum $GAD(v_h)$; 
		\State $D_h := D_{h-1}\cup \{v_h\}$;
		\State $C_h := C_{h-1}\cup N_{A_{h-1}}(v_h)\cup \overline{N}_{B_{h-1}}(v_h)$;
		\If{$A_{h-1} \neq \emptyset $}
		\State $A_h := A_{h-1}\setminus N_{A_{h-1}}(v_h)$;
		\EndIf
		\If{$B_{h-1} \neq \emptyset $}
		\State $B_h := B_{h-1}\setminus \overline{N}_{B_{h-1}}(v_h)$;
		\EndIf
		\EndWhile
		
	\end{algorithmic}
	
\end{algorithm*}

\medskip

\begin{proposition}\label{obs_opt}
Algorithm H1 finds an optimal feasible solution if it halts at iteration $h$ 
with $|D_h| \leq 3$, $h=0,1$. 
\end{proposition}
\begin{proof}
We need to show that $|D_h|= \gamma_g(G)$ whenever $|D_h| \leq 3$.  
Note first that $|D_h|$ cannot be 1. Suppose $|D_h|=2$, i.e., $h=0$ and the set 
$D_0$ contains two vertices, say $u$ and $v$. Since the algorithm halted with
the set $D_0$, $A_0=B_0=\emptyset$ must hold, which yields equation (\ref{union})
and that $D_0$ is a global dominating set (Lemma \ref{obsD0}) and the 
lemma follows for $h=0$. If now $|D_h|=3$ ($h=1$),  then there
exists no pair of vertices $(u,v)$ satisfying equation \ref{union} (by the construction). 
Hence, $\gamma_g(G)\ge 3$. At the same time, since the algorithm has halted at 
iteration 1, $A_1 = B_1 = \emptyset$ and by Lemma \ref{lema1}, $C_1 = V \setminus D_1$. 
Hence, $D_1$ is a feasible solution and the lemma is proved. 
\end{proof}

\begin{figure}[H]
	\centering
	\includegraphics[width=0.80\linewidth]{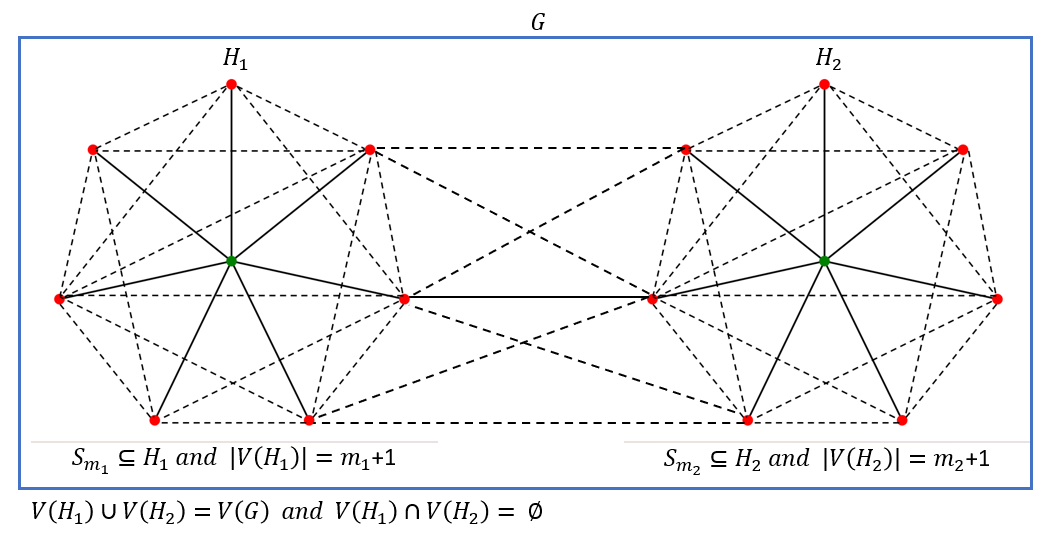}
	\caption{A graph $G$}
	\label{ejemplo2}
\end{figure}

Now we define a particular family of graphs. A connected graph $G=(V,E)$ from this 
family of order $n$ is composed of two sub-graphs $H_1$ and $H_2$ of order $m_i+1$ 
each containing the star graph $S_{m_i}$ as a sub-graph with
	$V(H_1)\cup V(H_2) = V(G)$ and $V(H_1) \cap V(H_2) =\emptyset$ and with 
	center  $v_i$ ($i=1,2$),  a vertex with maximum degree 	$m_i$ in  graph 
	 $S_{m_i}$. In Figure \ref{ejemplo2}  we illustrate such graph, where each dashed
	 line represents an edge that may exist or not. Note that each $H_i$ can also be
	 a complete graph.  
	
\begin{theorem}	
	Algorithm H1 finds an optimal solution for any graph $G$ from the above specified family. 	
\end{theorem}
\begin{proof}
	Given graph $G$, 
	in the first iteration, Algorithm H1 will select the two vertices $\{v_1, v_2\}$ 
	(marked in green in Figure \ref{ejemplo2}), such that 
	$|(N[v_1] \cup N[v_2]) \setminus (N[v_1]\cap N[v_2])|$ is maximum.
	We easily observe that 
	$(N[v_1] \cup N[v_2]) \setminus (N[v_1]\cap N[v_2]) = V \setminus \{v_1,v_2\}$
	and hence $A_0=B_0=\emptyset$.  Therefore, Algorithm H1 will return a global 
	dominating set 	$D=\{v_1,v_2\}$ at iteration $0$, 
	which is optimal by Lemma \ref{obsD0}.
\end{proof}

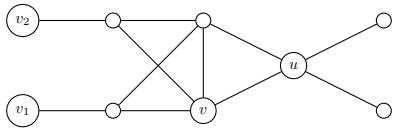
\begin{figure}[h!]
	\centering
	\begin{tikzpicture}[scale=.6, transform shape]

		\node [draw, shape=circle] (s1) at  (0,0) {};
		\node [draw, shape=circle] (s2) at  (0,2) {};
		\node [draw, shape=circle] (s3) at  (2,2) {};
		\node [draw, shape=circle] (s4) at  (2,0) {$v$};
		\node [draw, shape=circle] (s5) at  (4,1) {$u$};
		
		\node [draw, shape=circle] (s6) at  (6,0) {};
		\node [draw, shape=circle] (s7) at  (6,2) {};
		
		\node [draw, shape=circle] (s8) at  (-2,0) {$v_1$};
		\node [draw, shape=circle] (s9) at  (-2,2) {$v_2$};
		
		\draw(s1)--(s4)--(s3)--(s2)--(s4);
		\draw(s1)--(s3); \draw(s3)--(s5)--(s4); \draw(s6)--(s5)--(s7);
		\draw(s2)--(s9);
		\draw(s1)--(s8);

	\end{tikzpicture}
	\caption{A graph where the vertex set $\{u,v,v_1,v_2\}$ is a possible solution given by Algorithm \ref{alg_one}, however the vertex set $\{u,v_1,v_2\}$ is a global dominating set of minimum cardinality.}\label{fig-alg-1}
\end{figure}

In Figure \ref{fig-alg-1}, is shown an example where $|D_h|=4$, but the global domination number is 3.

The next proposition easily follows.

\begin{proposition}\label{lema_H1}
Algorithm H1 runs in at most $|A_0|+|B_0|$ iterations and hence its time complexity
is $O(n^3)$. 
\end{proposition}

\section{Algorithm H2}

Our next heuristic algorithm H2, at each iteration complements the current vertex 
set $D_h$ with two vertices, $v_h$ and $u_h$, defined (differently from the
previous section) as follows: $v_h$ is  a vertex dominating the
maximum number of yet non-dominated vertices in graph $G$, i.e., it has 
the maximum number of the adjacent vertices among those who do not yet 
have a neighbor in set $D_h$. Likewise, vertex $u_h$ is one that 
dominates the maximum number of yet non-dominated vertices in graph $\overline G$, 
i.e., it has the maximum number of non-adjacent vertices among those that 
are  not adjacent with a vertex in set $D_h$. 
In other words, $v_h\in V\setminus D_h$ is a vertex with  
$$N_{V\setminus N[D_h]}[v_h]=\max_{v\in V\setminus D_h} N_{V\setminus N[D_h]}[v].$$  
Likewise, $u_h\in V\setminus D_h$ is a vertex with
$$\overline{N}_{V\setminus \overline{N}[D_h]}[u_h]=\max_{v\in V\setminus D_h} \overline{N}_{V\setminus \overline{N}[D_h]}[v].$$
A detailed description  of the algorithm follows. 


\begin{algorithm*}
	\caption{H2}\label{alg_two}
	\fontsize{7pt}{.25cm}\selectfont
	\begin{algorithmic}
		\State Input: A graph $G(V,E)$.
		\State Output: A global dominating set $D_h$.
		
		\State $D_0 := \emptyset$;
		\State $C_0:= \emptyset$;  
		
		\{ iterative step \} 
		\State $h:= 0$; 
		\While { $D_h \cup C_h \neq V$ } 
		\State $h:= h + 1$;
		\State $v_h:=u_h:=NIL$;
		
		\If { $ D_h $ is not a dominant set of $ G $ }
		\State $v_h$ := a vertex in set $V \setminus D$ such that $N_{V\setminus N[D]}[v]$  be maximum with $\overline{N}_{V\setminus \overline{N}[D]}[v]$ of maximum cardinality;
		\EndIf 
		\State $D_h := D_{h-1} \cup \{v_h\}$;
		\If { $ D_h $ is not a dominant set of $\overline G $ }
		\State $u_h$ := a vertex in set $V \setminus D$ such that $\overline{N}_{V\setminus \overline{N}[D]}[u]$ be maximum with $N_{V\setminus N[D]}[u]$ of maximum cardinality; 
		\State $D_h := D_{h} \cup \{u_h\}$;
		\EndIf
		
		\State $C_h := N(D_h)\cap \overline{N}(D_h)$;
		\EndWhile
		
	\end{algorithmic}
	
\end{algorithm*}

\begin{figure}[h!]
	\centering
	\begin{tikzpicture}[scale=.4, transform shape]
		
		\node [draw, shape=circle] (a1) at  (0,0) {4};
		\node [draw, shape=circle] (a2) at  (4,0) {1};
		\node [draw, shape=circle] (a3) at  (5.5,3) {};
		\node [draw, shape=circle] (a4) at  (2,6) {};
		\node [draw, shape=circle] (a5) at  (-1.5,3) {2};
		
		\node [draw, shape=circle] (a6) at  (1,1) {3};
		\node [draw, shape=circle] (a7) at  (3,1) {};
		\node [draw, shape=circle] (a8) at  (4,3) {};
		\node [draw, shape=circle] (a9) at  (2,4.5) {};
		\node [draw, shape=circle] (a10) at  (0,3) {};
		
		\draw(a1)--(a2)--(a3)--(a4)--(a5)--(a1);
		\draw(a1)--(a6); \draw(a2)--(a7); \draw(a3)--(a8); \draw(a4)--(a9); \draw(a5)--(a10);
		\draw(a6)--(a9)--(a7)--(a10)--(a8)--(a6);
		
		\node [draw, shape=circle] (q1) at  (10,0) {};
		\node [draw, shape=circle] (q2) at  (14,0) {};
		\node [draw, shape=circle] (q3) at  (14,4) {};
		\node [draw, shape=circle] (q4) at  (10,4) {};
		
		\node [draw, shape=circle, fill=black] (q5) at  (8,2) {};
		\node [draw, shape=circle, fill=black] (q6) at  (16,2) {};
		\node [draw, shape=circle, fill=black] (q7) at  (12,6) {};
		\node [draw, shape=circle] (q8) at  (8,4.3) {};
		\node [draw, shape=circle] (q9) at  (16,4.3) {};
		
		\draw(q1)--(q2)--(q3)--(q4)--(q1)--(q3);
		\draw(q1)--(q5)--(q4); \draw(q2)--(q6)--(q3); \draw(q3)--(q7)--(q4);
		\draw(q5)--(q8)--(q4); \draw(q3)--(q9)--(q6); \draw(q8)--(q2);
		
		\node [font=\Large] at (2,-1) {$(a)$};
		\node [font=\Large] at (12,-1) {$(b)$};
		
		\node [font=\Large] at (14,4.5) {$v_1$};
		\node [font=\Large] at (7.5,2) {$v_2$};
		\node [font=\Large] at (10,-0.5) {$v_3$};
		\node [font=\Large] at (10,4.5) {$v_4$};
		
	\end{tikzpicture}
	\caption{(a) The Petersen graph, where $\gamma_g(G)=4$ and Algorithm H2 
	delivers an optimal feasible solution $\{1,2,3,4\}$; 
	(b) a graph where the black vertex set represents an optimal feasible solution,
	 and $\{v_1,v_2,v_3,v_4\}$ a global dominating set obtained by Algorithm H2.}
 	\label{fig-alg-2}
\end{figure}
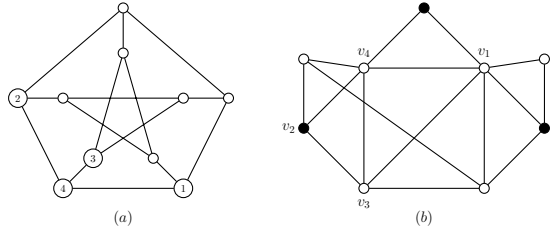

In Figure \ref{fig-alg-2} (a) and (b), we give examples when Algorithm H2 
obtains an optimal and non-optimal, respectively, feasible solutions. 

We may easily observe that the upper bounds $ U_2 $ and $ U_1 $ introduced in Section 2
(Theorems \ref{teorema5} and \ref{teorema6}) are attainable for example, for the complete graph 
with $n$ vertices $ K_n $ and the Petersen graph depicted in Figure \ref{fig-alg-2}(a), respectively. 
At the same time, for both types of graphs, Algorithm H2 returns an optimal solution (see 
again Figure \ref{fig-alg-2}(a) illustrating the Petersen graph and the optimal solution
generated by the algorithm for that graph). 

\begin{figure}[h!]
	\centering
	\begin{tikzpicture}[scale=.33, transform shape]
		
		\node [draw, shape=circle] (a1) at  (0,0) {};
		\node [draw, shape=circle] (a2) at  (4,0) {$v_1$};
		\node [draw, shape=circle] (a3) at  (5.5,3) {};
		\node [draw, shape=circle] (a4) at  (2,6) {};
		\node [draw, shape=circle] (a5) at  (-1.5,3) {$x_1$};
		
		\node [draw, shape=circle] (a6) at  (1,1) {$y_1$};
		\node [draw, shape=circle] (a7) at  (3,1) {};
		\node [draw, shape=circle] (a8) at  (4,3) {};
		\node [draw, shape=circle] (a9) at  (2,4.5) {};
		\node [draw, shape=circle] (a10) at  (0,3) {};
		
		\draw(a1)--(a2)--(a3)--(a4)--(a5)--(a1);
		\draw(a1)--(a6); \draw(a2)--(a7); \draw(a3)--(a8); \draw(a4)--(a9); \draw(a5)--(a10);
		\draw(a6)--(a9)--(a7)--(a10)--(a8)--(a6);
		
		
		\node [draw, shape=circle] (a12) at  (10,0) {};
		\node [draw, shape=circle] (a22) at  (14,0) {$v_2$};
		\node [draw, shape=circle] (a32) at  (15.5,3) {};
		\node [draw, shape=circle] (a42) at  (12,6) {};
		\node [draw, shape=circle] (a52) at  (8.5,3) {$x_2$};
		
		\node [draw, shape=circle] (a62) at  (11,1) {$y_2$};
		\node [draw, shape=circle] (a72) at  (13,1) {};
		\node [draw, shape=circle] (a82) at  (14,3) {};
		\node [draw, shape=circle] (a92) at  (12,4.5) {};
		\node [draw, shape=circle] (a102) at  (10,3) {};
		
		\draw(a12)--(a22)--(a32)--(a42)--(a52)--(a12);
		\draw(a12)--(a62); \draw(a22)--(a72); \draw(a32)--(a82); \draw(a42)--(a92); \draw(a52)--(a102);
		\draw(a62)--(a92)--(a72)--(a102)--(a82)--(a62);
		
		
		\node [draw, shape=circle] (a13) at  (20,0) {};
		\node [draw, shape=circle] (a23) at  (24,0) {$v_{n-1}$};
		\node [draw, shape=circle] (a33) at  (25.5,3) {};
		\node [draw, shape=circle] (a43) at  (22,6) {};
		\node [draw, shape=circle] (a53) at  (18.5,3) {$x_{n-1}$};
		
		\node [draw, shape=circle] (a63) at  (21,1) {$y_{n-1}$};
		\node [draw, shape=circle] (a73) at  (23,1) {};
		\node [draw, shape=circle] (a83) at  (24,3) {};
		\node [draw, shape=circle] (a93) at  (22,4.5) {};
		\node [draw, shape=circle] (a103) at  (20,3) {};
		
		\draw(a13)--(a23)--(a33)--(a43)--(a53)--(a13);
		\draw(a13)--(a63); \draw(a23)--(a73); \draw(a33)--(a83); \draw(a43)--(a93); \draw(a53)--(a103);
		\draw(a63)--(a93)--(a73)--(a103)--(a83)--(a63);
		
		
		\node [draw, shape=circle] (a14) at  (30,0) {};
		\node [draw, shape=circle] (a24) at  (34,0) {$v_{n}$};
		\node [draw, shape=circle] (a34) at  (35.5,3) {};
		\node [draw, shape=circle] (a44) at  (32,6) {};
		\node [draw, shape=circle] (a54) at  (28.5,3) {$x_{n}$};
		
		\node [draw, shape=circle] (a64) at  (31,1) {$y_{n}$};
		\node [draw, shape=circle] (a74) at  (33,1) {};
		\node [draw, shape=circle] (a84) at  (34,3) {};
		\node [draw, shape=circle] (a94) at  (32,4.5) {};
		\node [draw, shape=circle] (a104) at  (30,3) {};
		
		\draw(a14)--(a24)--(a34)--(a44)--(a54)--(a14);
		\draw(a14)--(a64); \draw(a24)--(a74); \draw(a34)--(a84); \draw(a44)--(a94); \draw(a54)--(a104);
		\draw(a64)--(a94)--(a74)--(a104)--(a84)--(a64);
		
		
		\node [draw, shape=circle] (p1) at  (4,-2) {};
		\node [draw, shape=circle] (p2) at  (14,-2) {};
		\node [font=\Large] (p) at  (19,-2) {($ \ldots $)};
		\node [draw, shape=circle] (p3) at  (24,-2) {};
		\node [draw, shape=circle] (p4) at  (34,-2) {};
		\draw(p1)--(p2);\draw(p2)--(p); \draw(p)--(p3); \draw(p3)--(p4);
		\draw(p1)--(a2); \draw(p2)--(a22); \draw(p3)--(a23); \draw(p4)--(a24);
		
		\draw [dashed] (p1)to [out=330,in=210,looseness=0.7](p4);
		\draw [dashed] (p1)to [out=340,in=200,looseness=0.8](p3);
		\draw [dashed] (p2)to [out=340,in=200,looseness=0.8](p4);

	\end{tikzpicture}
	\caption{Graph $G^\prime = G\circ_v H$, where $H$ is a Petersen graph.}
	\label{fig-alg-2a}
\end{figure}
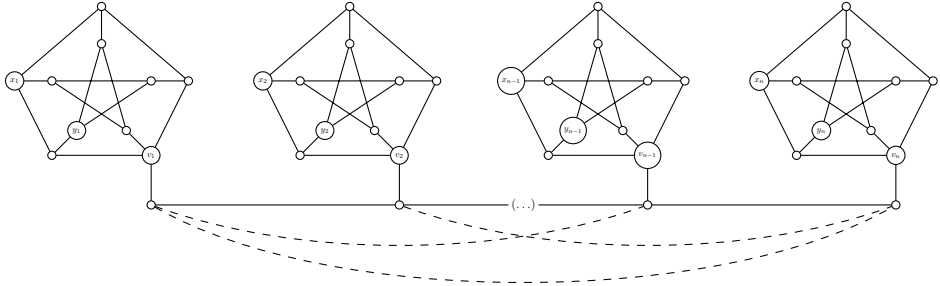

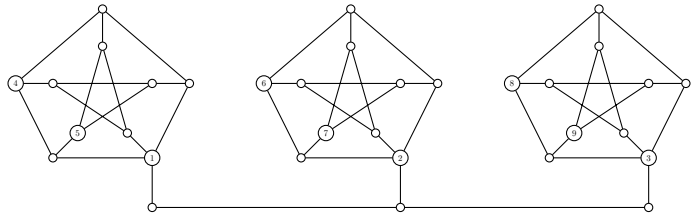
\begin{figure}[H]
	\centering
	\begin{tikzpicture}[scale=.33, transform shape]
		
		\node [draw, shape=circle] (a1) at  (0,0) {};
		\node [draw, shape=circle] (a2) at  (4,0) {1};
		\node [draw, shape=circle] (a3) at  (5.5,3) {};
		\node [draw, shape=circle] (a4) at  (2,6) {};
		\node [draw, shape=circle] (a5) at  (-1.5,3) {4};
		
		\node [draw, shape=circle] (a6) at  (1,1) {5};
		\node [draw, shape=circle] (a7) at  (3,1) {};
		\node [draw, shape=circle] (a8) at  (4,3) {};
		\node [draw, shape=circle] (a9) at  (2,4.5) {};
		\node [draw, shape=circle] (a10) at  (0,3) {};
		
		\draw(a1)--(a2)--(a3)--(a4)--(a5)--(a1);
		\draw(a1)--(a6); \draw(a2)--(a7); \draw(a3)--(a8); \draw(a4)--(a9); \draw(a5)--(a10);
		\draw(a6)--(a9)--(a7)--(a10)--(a8)--(a6);
		
		
		\node [draw, shape=circle] (a12) at  (10,0) {};
		\node [draw, shape=circle] (a22) at  (14,0) {2};
		\node [draw, shape=circle] (a32) at  (15.5,3) {};
		\node [draw, shape=circle] (a42) at  (12,6) {};
		\node [draw, shape=circle] (a52) at  (8.5,3) {6};
		
		\node [draw, shape=circle] (a62) at  (11,1) {7};
		\node [draw, shape=circle] (a72) at  (13,1) {};
		\node [draw, shape=circle] (a82) at  (14,3) {};
		\node [draw, shape=circle] (a92) at  (12,4.5) {};
		\node [draw, shape=circle] (a102) at  (10,3) {};
		
		\draw(a12)--(a22)--(a32)--(a42)--(a52)--(a12);
		\draw(a12)--(a62); \draw(a22)--(a72); \draw(a32)--(a82); \draw(a42)--(a92); \draw(a52)--(a102);
		\draw(a62)--(a92)--(a72)--(a102)--(a82)--(a62);
		
		
		\node [draw, shape=circle] (a13) at  (20,0) {};
		\node [draw, shape=circle] (a23) at  (24,0) {3};
		\node [draw, shape=circle] (a33) at  (25.5,3) {};
		\node [draw, shape=circle] (a43) at  (22,6) {};
		\node [draw, shape=circle] (a53) at  (18.5,3) {8};
		
		\node [draw, shape=circle] (a63) at  (21,1) {9};
		\node [draw, shape=circle] (a73) at  (23,1) {};
		\node [draw, shape=circle] (a83) at  (24,3) {};
		\node [draw, shape=circle] (a93) at  (22,4.5) {};
		\node [draw, shape=circle] (a103) at  (20,3) {};
		
		\draw(a13)--(a23)--(a33)--(a43)--(a53)--(a13);
		\draw(a13)--(a63); \draw(a23)--(a73); \draw(a33)--(a83); \draw(a43)--(a93); \draw(a53)--(a103);
		\draw(a63)--(a93)--(a73)--(a103)--(a83)--(a63);
		
		\node [draw, shape=circle] (p1) at  (4,-2) {};
		\node [draw, shape=circle] (p2) at  (14,-2) {};
		\node [draw, shape=circle] (p3) at  (24,-2) {};
		\draw(p1)--(p2); \draw(p2)--(p3);
		\draw(p1)--(a2); \draw(p2)--(a22); \draw(p3)--(a23);

	\end{tikzpicture}
	\caption{Graph $G^\prime = P_3\circ_v H$, where $H$ is a Petersen graph. The global domination number is 9 and Algorithm H2 obtain an optimal solution choosing the vertices in the shown order}
	\label{fig-alg-2b}
\end{figure}

\begin{figure}[H]
	\centering
	\includegraphics[width=0.8\linewidth]{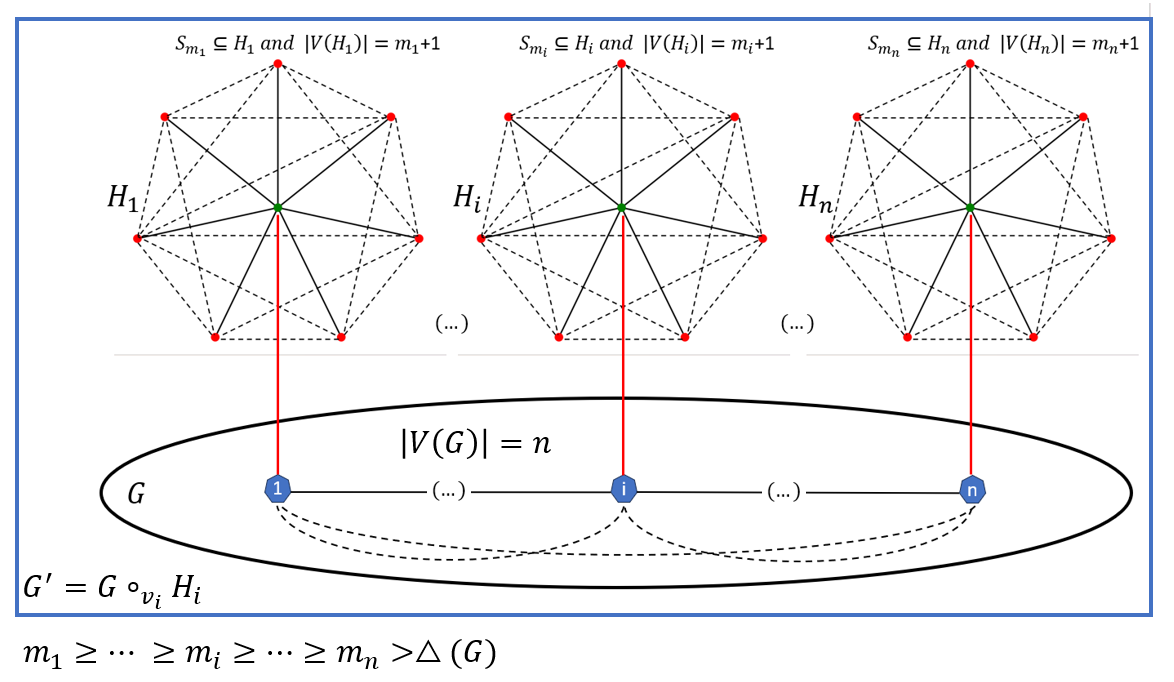}
	\caption{Graph $G^\prime = G\circ_{v_i} H_i$. The global domination number is $n$ and Algorithm H2 obtain an optimal solution choosing the green vertices.}
	\label{ejemplo}
\end{figure}

Next, we define a wider family of graphs for which the heuristic is optimal. 
Let $G$ with $|V(G)|=n$  and $H_i$, $i=1,2,\dots,n$ be connected graphs, all of the 
order no smaller than two, and let $G^\prime = G\circ_{v_i} H_i$ be the graph 
composed of these $n+1$ graphs, in which every vertex from graph $G$ is associated 
with a specially determined vertex $v_i$ of graph $H_i$ (as we will see, we will 
choose vertex $v_i$ from a $\gamma(H_i)$-set). Figure \ref{fig-alg-2a} illustrates 
an example of such a compouned graph, where every $H_i$ is the Petersen graph,
where in Figure \ref{fig-alg-2b} $G$ is the line graph $P_n$. In Figure  \ref{ejemplo} 
we illustrate another such graph, where $H_i$s are different. 

\begin{proposition}\label{prop1}
	Given graph $G^\prime = G\circ_{v_i} H_i$ with $|V(G)|=n\geq2$ and 
	$ v_i \in \gamma(H_i)$-set, 	$\gamma_g(G^\prime)$-set is 
	$\cup_{i=1}^{n}\gamma(H_i)$-set.  
\end{proposition}

\begin{proof}
	As it is easy to see, $n\geq 2$ implies that any two $v_i$-type vertices form 
	a dominating set of graph 	$ \overline{G^\prime} $,  whereas 
	$\{ v_i| i=\overline{1,n}\}$ form a dominating set of graph $ G $ since 
	$ v_i \in \gamma(H_i)$-set. Hence, $ \bigcup \limits_{i=1}^n \gamma(H_i)$-set 
	is 	a global dominating set of $ G^\prime $.       	
\end{proof}

For example, if  $ H_i$s are the Petersen graphs, then by above proposition,
$\gamma_g(G^\prime)=3n$   (see Figure \ref{fig-alg-2a}). If, in addition $G$ is $P_n$ 
or $C_n$ (the cycle graph of $n$ vertices), then Algorithm H2 will find an optimal 
solution for graph $G^\prime$ with $\gamma(G^\prime)=3\gamma(H_i)=9$, see Figure 
\ref{fig-alg-2b}. Note that since $\gamma_g(G^\prime)$-set and $\gamma(G^\prime)$-set 
are identical, the lower bound from Theorem \ref{teorema4} is attained. 

The next theorem deals with a more general family of graphs: 

\begin{theorem}
Let $H_i$, $i=\overline{1,n}$, be a connected graph of order $m_i+1$ 
containing the star graph $S_{m_i}$ as a subgraph, such that 
$m_1 \geq m_2 \geq \ldots \geq m_i \geq \ldots \geq m_{n-1} \geq m_n > \Delta(G)$.
Let, further, $v_i$ be the center of graph $S_{m_i}$ (a vertex with maximum degree 
$m_i$ in that graph), and let $G$ be a connected graph of order $n$. Then Algorithm 
H2 finds an optimal solution for graph $G^\prime = G\circ_{v_i} H_i$.	
\end{theorem}
\begin{proof}
Clearly, $\{v_i\}$ is a dominating set of graph $H_i$ and $\gamma_g(G^\prime)$-set
is $\cup_{i=1}^n \{v_i\}$ (see Proposition \ref{prop1} and Figure \ref{ejemplo}). 
Since $\Delta(G)$ is less than all $m_i$, Algorithm H2 in the first iteration 
will select the two vertices of type $v_i$ with the highest degrees, that is, 
$\{v_1, v_2\}$, that forms a dominating set of graph $ \overline{G^\prime}$. In 
every following iteration $h=2,\dots,n-1$, Algorithm H2 adds vertex  $v_{i+1}$ to 
the current solution and returns a global dominating set 
$D = \{v_i | i = \overline{i,n}\}$ at iteration $n-1$, which is optimal by 
Proposition \ref{prop1}.
\end{proof}

\section{Algorithm H3}

In our last Algorithm H3, the vertex sets $D_h$, $ A_h $, $ B_h $ and $ C_h $
and the initial iteration 0 with vertices $u$ and $v$ are defined as in Algorithm 
H1. 
The interactive parts of Algorithms H1 and H2 are combined as follows. 
If the condition $A_{h-1}=B_{h-1}=\emptyset$ (see Observation \ref{obsD0}) is 
satisfied, then the algorithm halts at iteration $h\ge 1$. Otherwise, it 
determines vertices $v_h,\; v_h^\prime,\; u_h$:
  
\begin{itemize}
	\item $v_h$ is a vertex in $V\setminus D_{h-1}$ with the maximum global 
	active degree; 	
	\item $v_h^\prime\in V\setminus D_{h-1}$ is a vertex with  
	$N_{V\setminus N[D_{h-1}]}[v_h^\prime]=\max_{v\in V\setminus D_{h-1}} N_{V\setminus N[D_{h-1}]}[v]$;  	
	\item $u_h\in V\setminus D_{h-1}$ is a vertex with
	$\overline{N}_{V\setminus \overline{N}[D_{h-1}]}[v_h]=\max_{v\in V\setminus D_{h-1}} \overline{N}_{V\setminus \overline{N}[D_{h-1}]}[v].$ 
\end{itemize}

Let $$GAD(v_h^\prime,u_h) = |N_{A_{h-1}}[\{v_h^\prime,u_h\}] \cup 
\overline{N}_{B_{h-1}}[\{v_h^\prime,u_h\}]|.$$ 
If $GAD(v_h) \geq GAD(v_h^\prime,u_h)/2$, then we let $D_h:= D_{h-1} \cup \{v_h\}$;
otherwise, we let $D_h:= D_{h-1} \cup \{v_h^\prime,u_h\}$. If $A_{h-1}=\emptyset$, 
vertex $ v_h^\prime $ is not defined, and if $ B_{h-1} =\emptyset $, vertex 
$ u_h $ is not defined.  The sets $A_h$, $B_h$ and $C_h$ are updated 
correspondingly. A detailed description is given below. 

\begin{algorithm*}[h!]
	\caption{H3}\label{alg_3}
	\fontsize{7pt}{.25cm}\selectfont
	\begin{algorithmic}
		\State Input: A graph $G(V,E)$.
		\State Output: A global dominating set $D_h$.
		\State $h := 0$; 
		\State $\{u,v\}$ := two vertex with the largest possible cardinality of the set \\ $(N[u]\cup N[v])\setminus (N[u]\cap N[v])$;
		\State $D_0:= \{u,v\}$;  
		\State $C_0:= (N[u]\cup N[v])\setminus (N[u]\cap N[v])$;  
		\State $A_0:= V\setminus (N[u]\cup N[v])$; 
		\State $B_0:= V\setminus (D_0\cup C_0 \cup A_0)$;  
		
		\{ iterative step \} 
		
		\While{ $A_h \neq \emptyset \vee B_h \neq \emptyset$ } 
		\State $h := h+1$; 
		\State $v_h$ := a vertex in the set $V \setminus D_{h-1}$ with the maximum $GAD(v_h)$; 
		
		\If { $ A_h \neq \emptyset $ }
		\State $v_h^\prime$ := a vertex in the set $V \setminus D_{h-1}$ such that $N_{V\setminus N[D_{h-1}]}[v_h^\prime]$  be maximum; 
		\EndIf
		
		\If { $ B_h \neq \emptyset $ }
		\State $u_h$ := a vertex in the set $V \setminus D_{h-1}$ such that $\overline{N}_{V\setminus \overline{N}[D_{h-1}]}[u_h]$ be maximum; 
		\EndIf
		
		\If { $GAD(v_h) \geq GAD(v_h^\prime,u_h)/2$}
		\State $D_h := D_{h-1}\cup \{v_h\}$;
		\State $C_h := C_{h-1}\cup N_{A_{h-1}}(v_h)\cup \overline{N}_{B_{h-1}}(v_h)$;
		\State $A_h := A_{h-1}\setminus N_{A_{h-1}}(v_h)$;
		\State $B_h := B_{h-1}\setminus \overline{N}_{B_{h-1}}(v_h)$;
		\Else
		\State $D_h := D_{h-1}\cup \{v_h^\prime,u_h\}$;
		\State $C_h := C_{h-1}\cup N_{A_{h-1}}(v_h^\prime,u_h)\cup \overline{N}_{B_{h-1}}(v_h^\prime,u_h)$;
		\State $A_h := A_{h-1}\setminus N_{A_{h-1}}(v_h^\prime,u_h)$;
		\State $B_h := B_{h-1}\setminus \overline{N}_{B_{h-1}}(v_h^\prime,u_h)$; 
		\EndIf 
		
		\EndWhile
		
	\end{algorithmic}
\end{algorithm*}



\section{Procedure Purify}

Purification procedures have been successfully used for the reduction of the set of 
vertices in a given dominating set. Below we apply similar kind of a reduction to
any given global dominating set. We need the following definition.

Let $D^{*}=\{v_1, ..., v_p\}$ be a formed global dominating set, the vertices being 
included in this particular order into that set. We define the purified set $D$ as
follows. Initially,  $D:= D^{*} $; iteratively, for $i = p, ..., 1$: if $pn(v_i,D,G)=\emptyset$ and $pn(v_i,D,\overline G)=\emptyset$ then we let 
$D:= D \setminus \{v_i\}$, i.e. we purify vertex $v_i$.

The so formed set $D$ is a minimal global dominating set for graph $G$ since, by
the construction, by eliminating any vertex, the resultant set will not 
globally dominate the  private neighbors of that vertex. 
It is easy to see that \textit{Procedure Purify} can be implemented in time 
$O(pn^2)=O(n^3)$. 

\section{Experimental Results} \label{sec_er}

In this section we describe our computation experiments. We implemented our algorithms in 
C++ using Windows 10 operative system for 64 bits on a personal computer with Intel Core i7-9750H (2.6 GHz) and 16 GB of RAM DDR4. We tested our algorithms for the benchmark 
problem instances from \cite{Parra2022}, 2284 instances in total. First, we describe
the results concerning our exact algorithms and then we deal with our heuristics. 
We analyze the performance of these heuristics based, in particular, on the optimal 
solutions obtained by our exact methods and known upper and lower bounds.  

\begin{figure}[H]
	\centering
	\begin{subfigure}[b]{0.48\linewidth}
		\includegraphics[width=\linewidth]{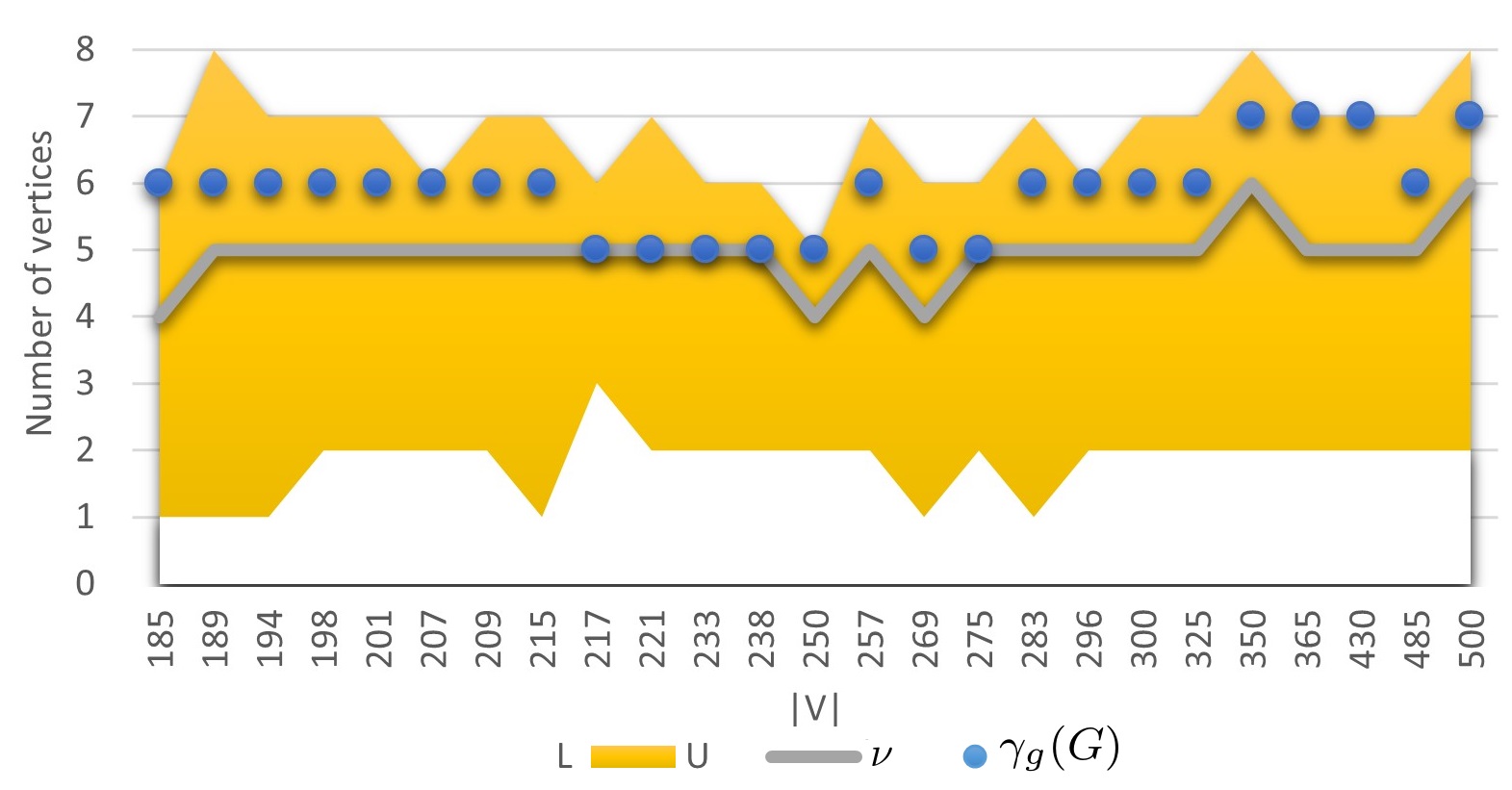}
		\caption{Iteration 1}
		\label{figura1_1}
	\end{subfigure}
	\begin{subfigure}[b]{0.48\linewidth}
		\includegraphics[width=\linewidth]{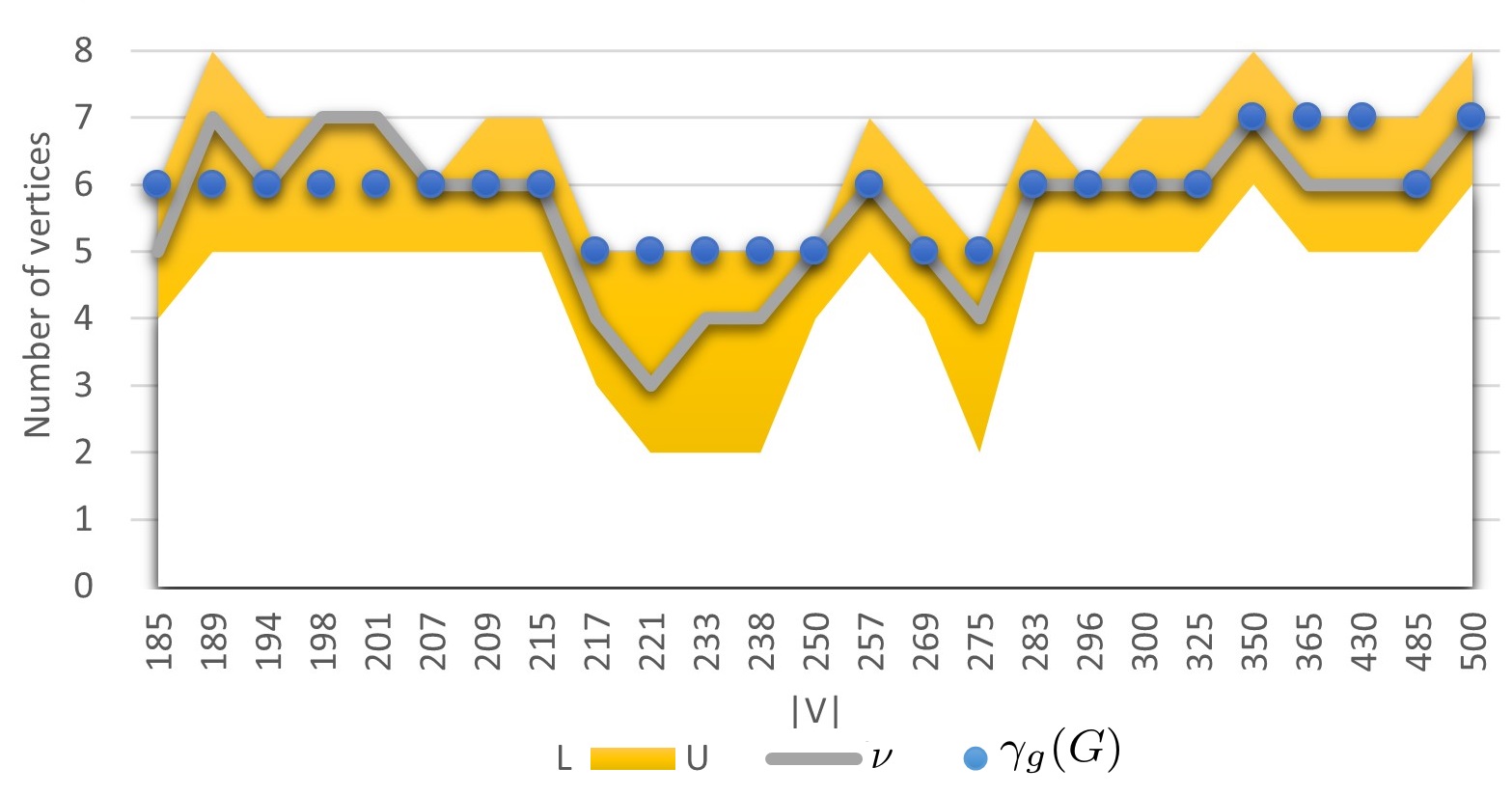}
		\caption{Iteration 2}
		\label{figura1_2}
	\end{subfigure}
	\begin{subfigure}[b]{0.48\linewidth}
		\includegraphics[width=\linewidth]{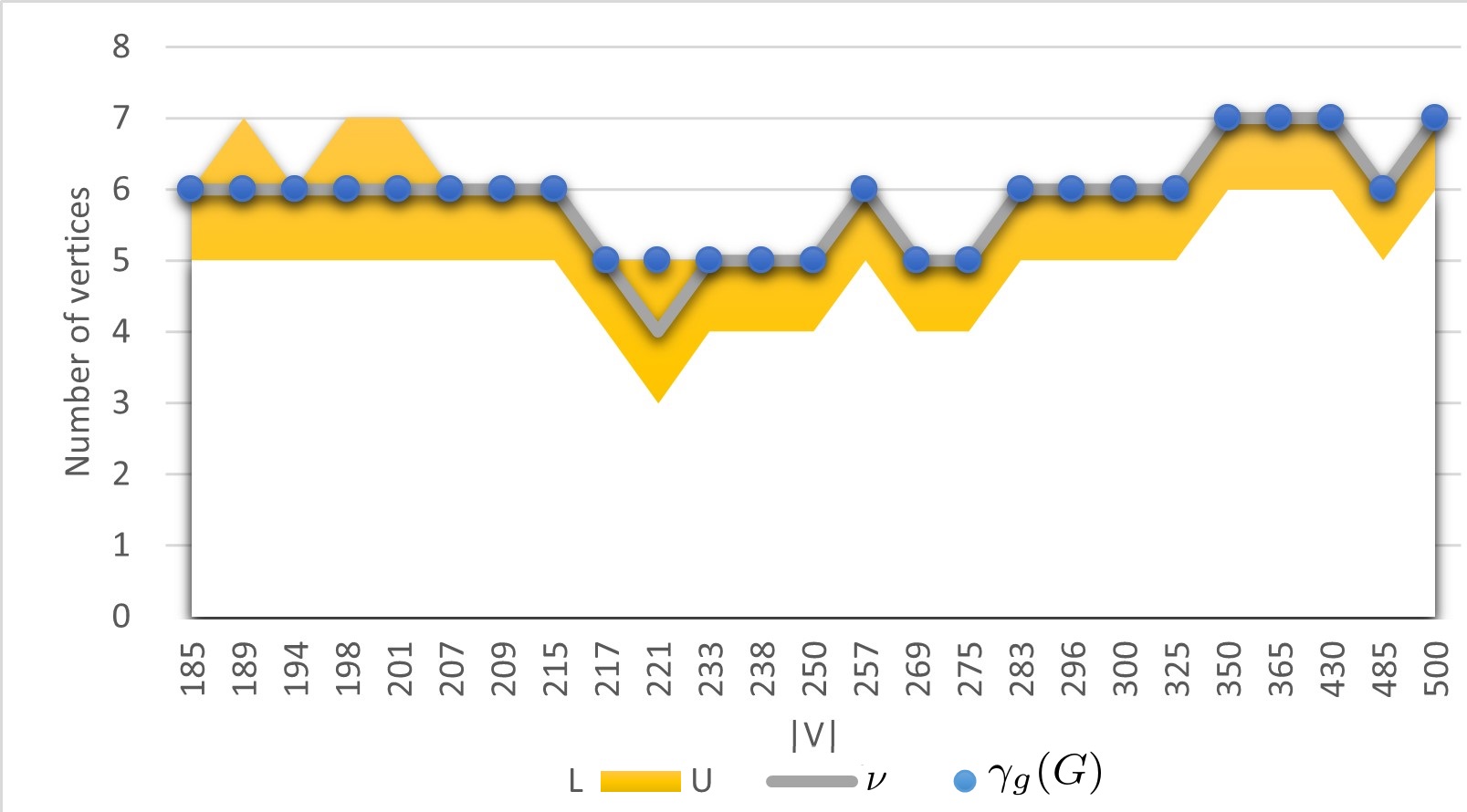}
		\caption{Iteration 3}
		\label{figura1_3}
	\end{subfigure}
	\begin{subfigure}[b]{0.48\linewidth}
		\includegraphics[width=\linewidth]{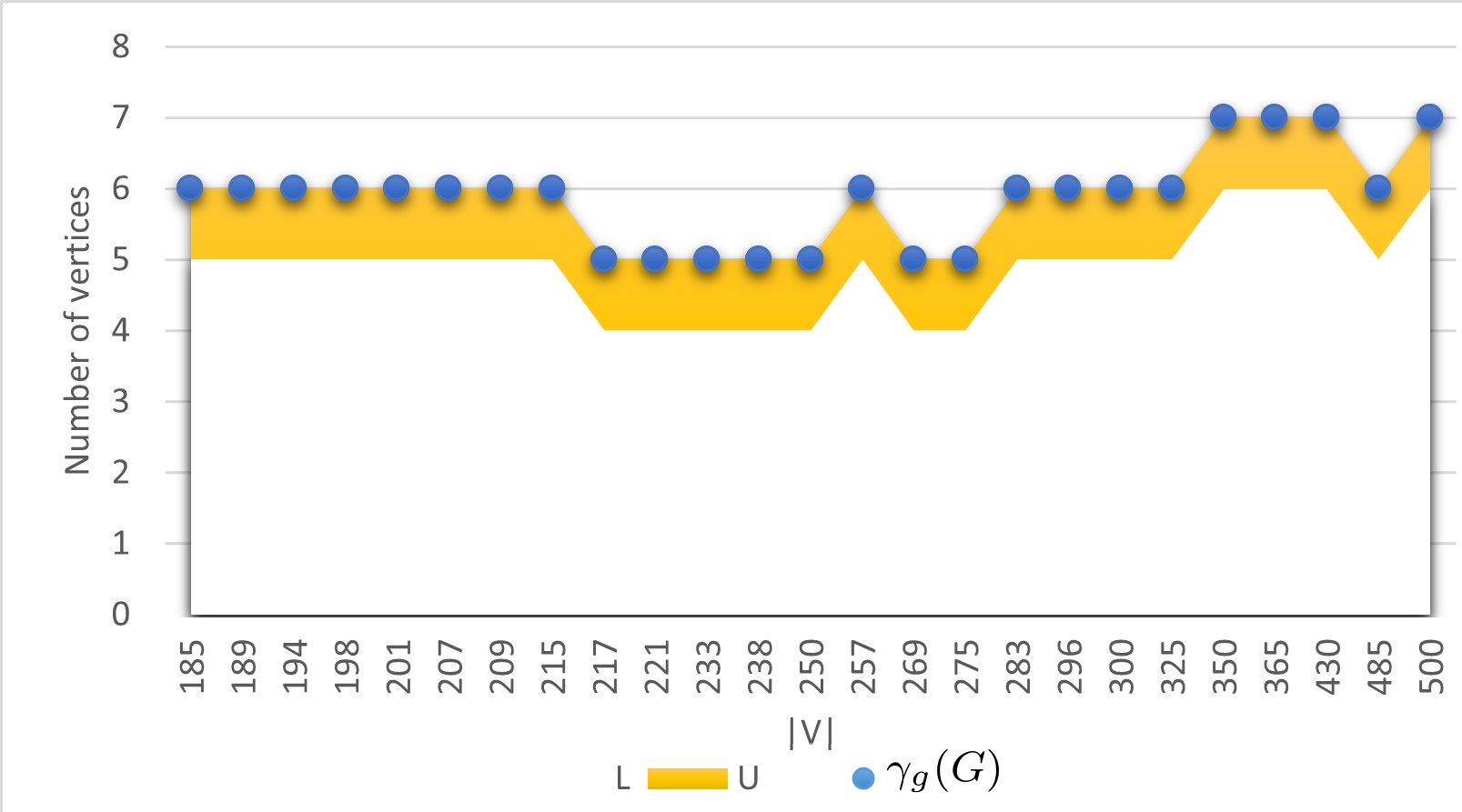}
		\caption{Iteration 4}
		\label{figura1_4}
	\end{subfigure}
	\caption{Vertical lines represent lower and upper bounds and the cardinalities 
	of the corresponding solutions, 
	and horizontal lines represent the size of the corresponding instances.  
	Ranges for lower and upper bounds are delineated by yellow regions, 
	where the gray lines show the current value of $\nu$; blue circles represent
	 $ \gamma_g(G)$}
	\label{figura1}
\end{figure}

{ \bf The performance of our exact algorithms.} For 25 middle-sized 
middle-dense instances, in at most four iterations an optimal solution was found  
by both BGDS and the ILP method.
Figure \ref{figura1} illustrates how the lower and upper bounds were updated 
during the execution of the algorithm and how the cardinality of the formed dominating
sets was iteratively reduced. 

Table \ref{tabla8} presents the results for small-sized graphs with the density of approximately 0.2 for both exact algorithms. For these instances, ILP formulation gave
better results (due to small amount of variables in linear programs).

\begin{footnotesize}
	\begin{longtable}[c]{ccccccccccc}
		\hline
		\multirow{2}{*}{No.} &
		\multirow{2}{*}{$|V(G)|$} &
		\multirow{2}{*}{$|E(G)|$} &
		\multicolumn{2}{c}{Time}&
		\multicolumn{4}{c}{Lower Bounds} &
		\multirow{2}{*}{$\gamma_g(G)$} &
		\multirow{2}{*}{U} \\ \cline{4-9} 
		& &  & ILP & BGDS & $\frac{n}{\Delta(G)+1}$ & $\frac{d+1}{3}$ & $\frac{2r}{3}$ & $|Supp(G)|$ & & \\ 
		\hline
		\endhead			
		1  & 50  & 286  & 0.18 & 1.79 & 2 & 1  & 1  & 2  & 7 & 7  \\ 
		2  & 60  & 357  & 0.13 & 4.16 & 2 & 1  & 1  & 1  & 6 & 7  \\ 
		3  & 70  & 524  & 0.16 & 7.44 & 2 & 1  & 1  & 1  & 7 & 8  \\ 
		4  & 80  & 678  & 0.14 & 11.93 & 2 & 1 & 1  & 1  & 6 & 6  \\ 
		5  & 90  & 842  & 0.18 & 384.92& 3 & 1 & 1  & 1  & 8 & 9  \\ 
		6  & 100 & 1031 & 0.23 & 794.41& 2 & 1 & 1  & 2  & 7 & 7  \\ 
		7  & 101 & 1051 & 1.31 & 811.32& 3 & 1 & 1  & 1  & 8 & 8  \\ 
		8  & 106 & 1154 & 1.57 & 973.85& 2 & 1 & 1  & 1  & 7 & 8  \\ 
		9  & 113 & 1306 & 1.34 &1384.32& 3 & 1 & 1  & 1  & 7 & 7  \\ 
		10 & 117 & 1398 & 2.73 &3474.51& 3 & 1 & 1  & 2  & 8 & 9  \\ 
		11 & 121 & 1493 & 3.11 &2586.79& 3 & 1 & 1  & 1  & 7 & 7  \\
		\hline
		\caption{The results for the graphs with density $\approx 0.2$.}
		\label{tabla8}
	\end{longtable}
\end{footnotesize}

For middle-dense graphs, where exact algorithms succeeded to halt with an optimal solution, Algorithm BGDS showed better performance for the graphs with more than 230 vertices,
where ILP formulation gave better results for smaller sized graphs (see
Table \ref{tabla9} and Figure \ref{figura2} below). 

\begin{footnotesize}
	\begin{longtable}[c]{ccccccccccc}
		\hline
		\multirow{2}{*}{No.} &
		\multirow{2}{*}{$|V(G)|$} &
		\multirow{2}{*}{$|E(G)|$} &
		\multicolumn{2}{c}{Time}&
		\multicolumn{4}{c}{Lower Bounds} &
		\multirow{2}{*}{$\gamma_g(G)$} &
		\multirow{2}{*}{U} \\ \cline{4-9} 
		& &  & ILP & BGDS & $\frac{n}{\Delta(G)+1}$ & $\frac{d+1}{3}$ & $\frac{2r}{3}$ & $|Supp(G)|$ & & \\ 
		\hline			
		
		1 & 185 & 6849 & 3.25 & 23.63    & 1 & 1 & 1 & 1 & 6 & 6 \\ 
		2 & 189 & 7147 & 3.87 & 22.55   &  1 & 1 & 1 & 1 & 6 & 8     \\ 
		3 & 194 & 7529 & 4.09 & 28.29    & 1 & 1 & 1 & 1 & 6 & 7     \\ 
		4 & 198 & 7842 & 3.71 & 31.53   &  2 & 1 & 1 & 1 & 6 & 7  \\ 
		5 & 201 & 8081 & 4.82 & 31.03   &  2 & 1 & 1 & 2 & 6 & 7   \\ 
		6 & 207 & 8569 & 5.28 & 33.91   & 2 & 1 & 1 & 1 & 6 & 6   \\ 
		7 & 209 & 8735 & 7.41 & 35.65   & 1 & 2 & 1 & 1 & 6 & 7   \\ 
		8 & 215 & 9243 & 12.25& 41.79   & 1 & 1 & 1 & 1 & 6 & 7   \\ 
		9 & 217 & 9415 & 19.16& 32.18    & 1 & 1 & 1 & 3 & 5 & 6   \\ 
		10 & 221 & 9765& 24.73& 39.91  & 2 & 1 & 1 & 1 & 5 & 7  \\ 
		
		11 & 233 & 10852& 32.74  & 27.83   & 2 & 1 & 1 & 1 & 5 & 6   \\ 
		12 & 238 & 11322& 51.49  & 45.56    &  2 & 1 & 1 & 1 & 5 & 6  \\ 
		13 & 250 & 12491& 86.09  & 74.19   &  1 & 2 & 1 & 1 & 5 & 5  \\ 
		14 & 257 & 13199& 128.43 & 94.89    &  2 & 1 & 1 & 1 & 6 & 7   \\ 
		15 & 269 & 14459& 134.03 & 102.60   &  1 & 1 & 1 & 1 & 5 & 6  \\ 
		16 & 275 & 15111& 583.66 & 124.46   &  2 & 1 & 1 & 1 & 5 & 6   \\ 
		17 & 283 & 16002& 1166.80& 436.23    &  1 & 1 & 1 & 1 & 6 & 7   \\ 
		18 & 296 & 17505& 5837.52& 385.61   &  2 & 1 & 1 & 1 & 6 & 6  \\ 
		19 & 300 & 17981& 6418.32& 405.03   & 2 & 1 & 1 & 1 & 6 & 7   \\ 
		
		20 & 325 & 24896 & 7002.01 & 571.27   & 1 & 2 & 1 & 1 & 6 &  7  \\
		21 & 350 & 27495 & 7854.36 & 3057.62   & 2 & 2 & 1 & 1 & 7 & 8   \\
		22 & 365 & 31606 & 8153.24 & 6686.34   & 2 & 1 & 1 & 1 & 7 & 7   \\
		23 & 430 & 44216 & 10121.83& 7001.24   & 1 & 2 & 1 & 1 & 7 & 7   \\
		24 & 485 & 56536 & 11936.22& 1484.51   & 1 & 2 & 1 & 1 & 6 & 7   \\
		25 & 500 & 60158 & 12351.49& 4525.39   & 2 & 1 & 1 & 1 & 7 & 8   \\
		
		\hline
		\caption{The results for the graphs with density $\approx 0.5$.\label{tabla9}}
	\end{longtable}
\end{footnotesize}

\begin{figure}[H]
	\centering
	\includegraphics[width=0.7\linewidth]{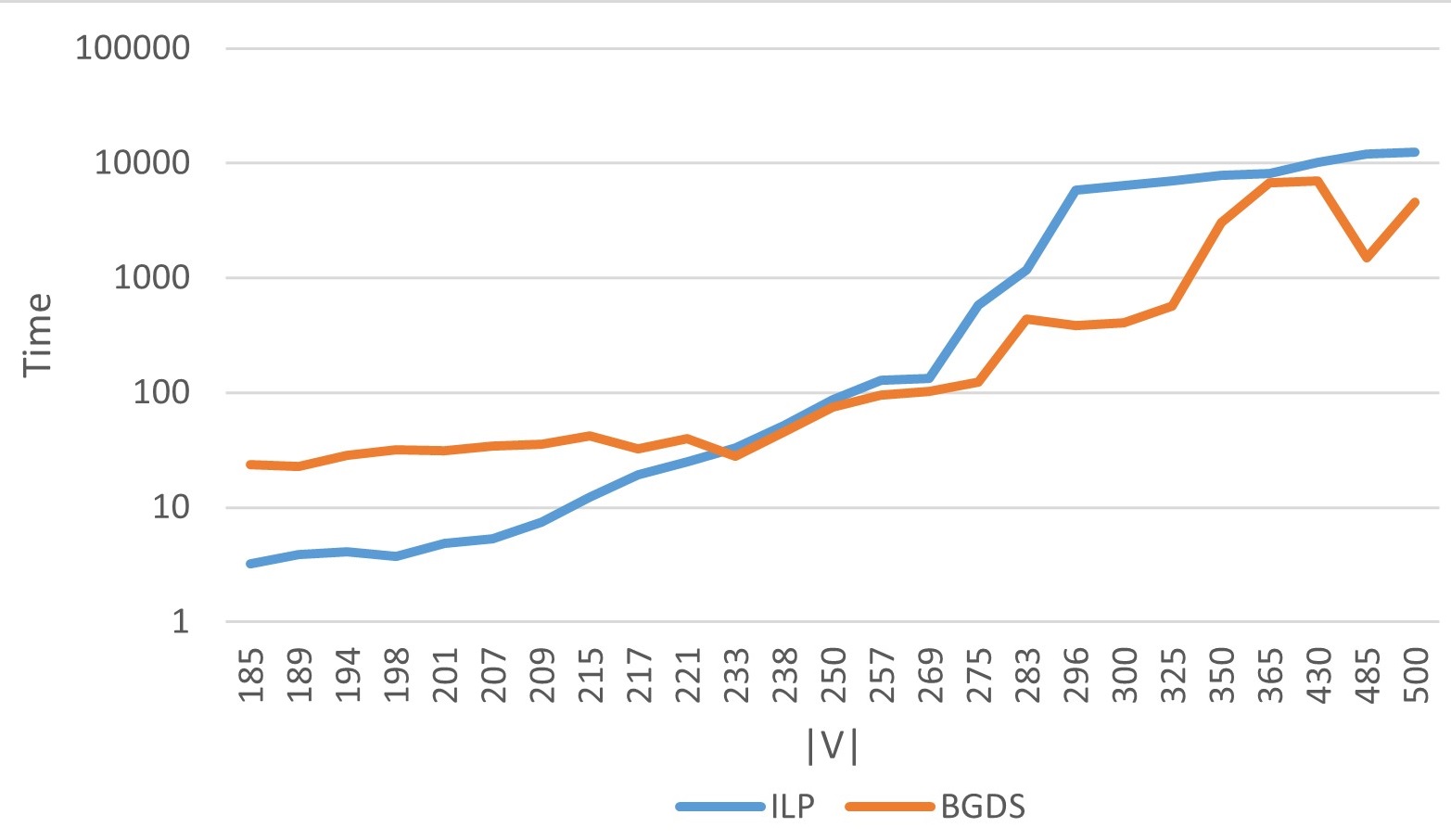}
	\caption{BGDS vs. ILP for the graphs with density $\approx 0.5$ (logarithmic scale was applied).}
	\label{figura2}
\end{figure}

\begin{figure}[H]
	\centering
	\includegraphics[width=0.70\linewidth]{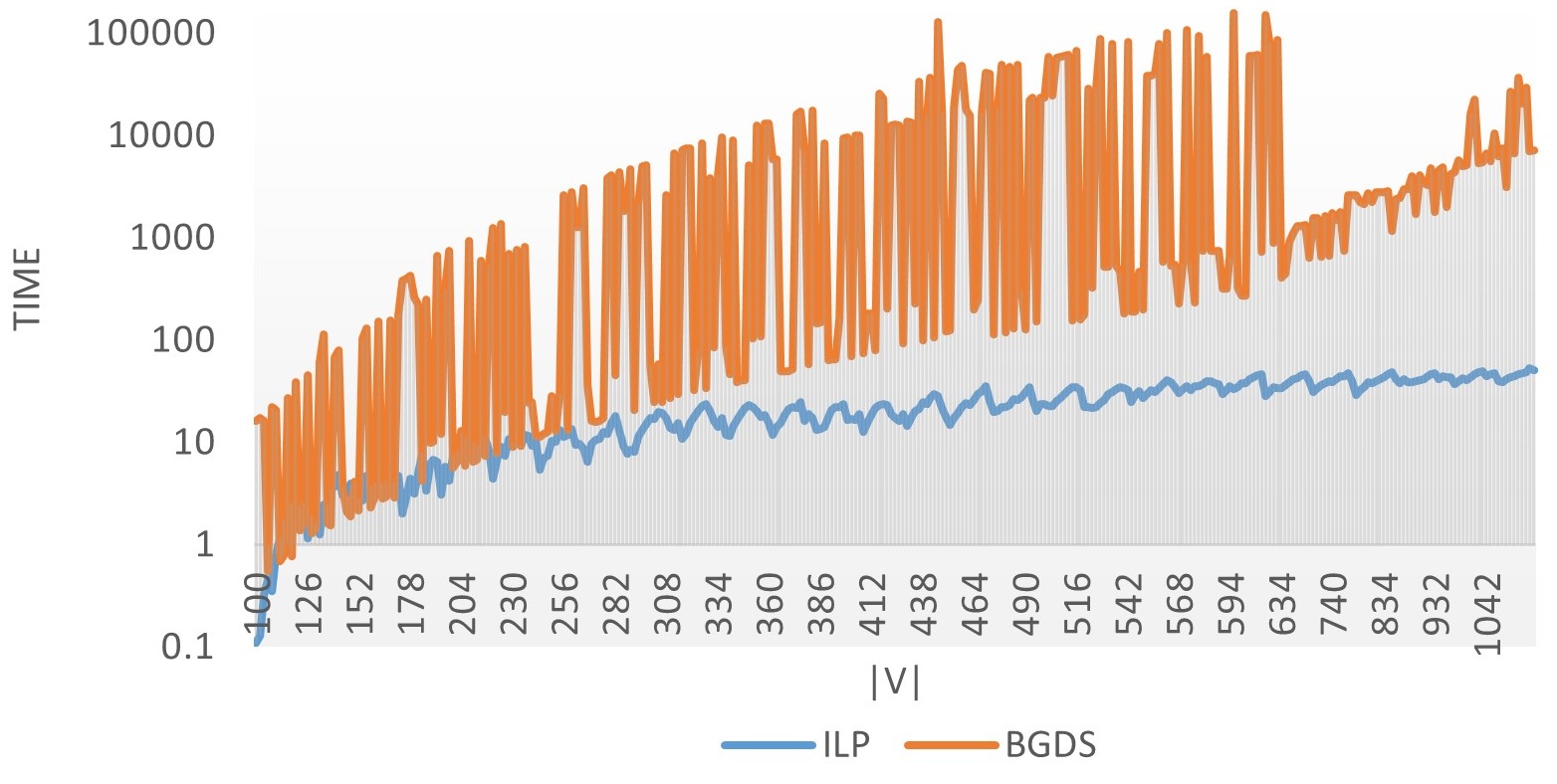}
	\caption{BGDS vs. ILP for dense graphs (logarithmic scale was applied).}
	\label{figura3}
\end{figure}

Table \ref{tabla3} presents results for larger  instances up to
1098 vertices with high density ($\approx 0.8$). Both exact algorithm completed within 
120 minutes for the largest instances. For these dense quasi complete graphs 
$\gamma_g(G)$ turned out to be very small (ranging from 4 to 7), where ILP formulation resulted in smaller execution times (see Figure \ref{figura3}).

\vspace{1cm}  

\begin{footnotesize}
	\begin{longtable}[c]{ccccccccccc}
		\hline
		\multirow{2}{*}{No.} &
		\multirow{2}{*}{$|V(G)|$} &
		\multirow{2}{*}{$|E(G)|$} &
		\multicolumn{2}{c}{Time}&
		\multicolumn{4}{c}{Lower Bounds} &
		\multirow{2}{*}{$\gamma_g(G)$} &
		\multirow{2}{*}{U} \\ \cline{4-9} 
		& &  & ILP & BGDS & $\frac{n}{\Delta(G)+1}$ & $\frac{d+1}{3}$ & $\frac{2r}{3}$ & $|Supp(G)|$ & & \\ 
		\hline
		
		1 &	250	&	23169	&	10.34	&	28.61	&	1	&	2	&	2	&	2	&	5	&	5	\\
		2 &	302	&	38543	&	16.89	&	24.97	&	1	&	2	&	2	&	2	&	4	&	5	\\
		3 &	352	&	46953	&	22.39	&	103.23	&	1	&	2	&	2	&	2	&	5	&	6	\\
		4 &	420	&	67432	&	23.09	&	205.13	&	1	&	2	&	2	&	2	&	5	&	5	\\
		5 &	474	&	97121	&	20.11	&	113.59	&	1	&	2	&	2	&	2	&	4	&	5	\\
		6 &	518	&	116381	&	32.56	&	160.37	&	1	&	2	&	2	&	2	&	4	&	5	\\
		7 &	580	&	130202	&	37.67	&	746.51	&	1	&	2	&	2	&	2	&	5	&	6	\\
		8 &	634	&	175537	&	33.93	&	410.16	&	1	&	2	&	2	&	2	&	4	&	5	\\
		9 &	712	&	222137	&	39.85	&	646.24	&	1	&	2	&	2	&	2	&	4	&	5	\\
		10&	746	&	244124	&	44.08	&	753.398	&	1	&	2	&	2	&	2	&	4	&	5	\\
		11&	816	&	260190	&	47.24	&	2604.72	&	1	&	2	&	2	&	2	&	5	&	6	\\
		12&	892	&	350495	&	41.63	&	3017.06	&	1	&	2	&	2	&	2	&	5	&	5	\\
		13&	932	&	382996	&	47.28	&	1805.78	&	1	&	2	&	2	&	2	&	4	&	5	\\
		14&	970	&	415180	&	43.83	&	4186.27	&	1	&	2	&	2	&	2	&	5	&	5	\\
		15&	972	&	416905	&	37.55	&	4360.8	&	1	&	2	&	2	&	2	&	5	&	5	\\
		
		16&	1014	&	454116	&	40.84	&	5136.9	&	1	&	2	&	2	&	2	&	5	&	5	\\
		17&	1036	&	474219	&	48.64	&	5341.2	&	1	&	2	&	2	&	2	&	5	&	5	\\
		18&	1042	&	426451	&	44.52	&	6742.94	&	1	&	2	&	2	&	2	&	5	&	6	\\
		19&	1064	&	500420	&	40.21	&	6224.74	&	1	&	2	&	2	&	2	&	5	&	5	\\
		20&	1068	&	504205	&	41.84	&	3173.52	&	1	&	2	&	2	&	2	&	4	&	5	\\
		21&	1080	&	515722	&	44.24	&	6698.43	&	1	&	2	&	2	&	2	&	5	&	5	\\
		22&	1096	&	531243	&	52.96	&	7010.88	&	1	&	2	&	2	&	2	&	5	&	5	\\
		23&	1098	&	533220	&	51.51	&	7246.31	&	1	&	2	&	2	&	2	&	5	&	5	\\
		
		\hline
		\caption{Results for the graphs with density $\approx 0.8$}
		\label{tabla3}
	\end{longtable}
\end{footnotesize}

\begin{figure}[H]
	\centering
	\includegraphics[width=0.80\linewidth]{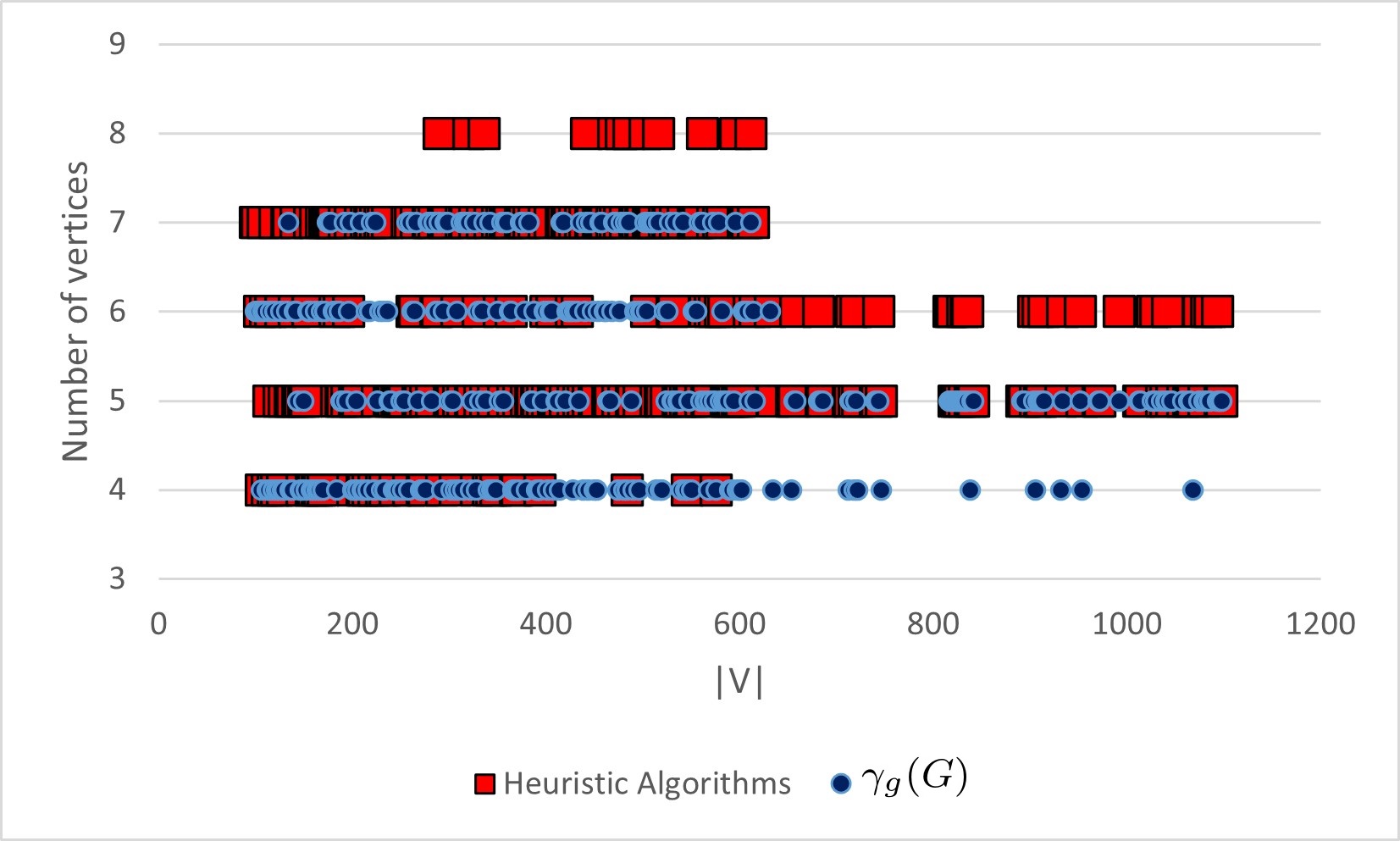}
	\caption{$\gamma_g(G)$ vs. approximation results.}
	\label{figura4}
\end{figure}

{ \bf The performance of the heuristic algorithms.}
In Figure \ref{figura4} we compare the quality of the best solution generated by 
	our heuristic algorithms  vs optimal values obtained by our exact algorithms 
	for the large-sized instances with up to 1098 vertices. In this figure, blue
	nodes represent optimal solutions, and red squares in black frames represent
	a best solution obtained by our heuristics. As we can see, 
	the heuristics found an optimal solution in $52.03\%$ of the instances, whereas, 
	among the instances where the optimum was not found, the average approximation error 
	was $1.07$.

The above reported results of our heuristics include the purification stage. 
In fact, the purification process turned out to be efficient, in practice, as
illustrated in Tables \ref{table1} and \ref{table6}, in which the values of 
the corresponding lower and upper bounds  $ L $ and $ U $ are also shown; $|D|$ and
$|D^*|$, respectively,  stand for the number of vertices in the global dominant set 
obtained before and after, respectively, purification.

\begin{scriptsize}
	\begin{longtable}[c]{ccccccccccccc} 
		\hline
		\multirow{2}{*}{No.} & \multirow{2}{*}{$|V|$} & \multirow{2}{*}{$|E|$} & \multirow{2}{*}{L} & \multicolumn{4}{c}{Algorithm H1} & \multicolumn{3}{c}{Algorithm H2} &  & \multirow{2}{*}{U} \\ \cline{5-12}
		&  &  &  & Time(s) & $|D_{H1}|$ & $|D_{H1}^*|$ & \% Purified & Time(s) & $|D_{H2}|$ & $|D_{H2}^*|$ & \% Purified &  \\ \hline
		\endhead
		\hline
		\endfoot
		\endlastfoot
		1 & 7300 & 7311 & 2663 & 2794.67 & 3344 & 2706 & 19.08 & 567.73 & 3161 & 2706 & 14.39 & 6843 \\
		2 & 7350 & 7474 & 2621 & 2839.76 & 3365 & 2726 & 18.99 & 578.91 & 3177 & 2722 & 14.32 & 6890 \\
		3 & 7400 & 7474 & 2677 & 2949.72 & 3392 & 2743 & 19.13 & 591.42 & 3202 & 2740 & 14.43 & 6937 \\
		4 & 7450 & 7497 & 2695 & 2996.26 & 3399 & 2756 & 18.92 & 607.24 & 3218 & 2755 & 14.39 & 6984 \\
		5 & 7500 & 7535 & 2762 & 3077.71 & 3427 & 2822 & 17.65 & 615.30 & 3270 & 2822 & 13.70 & 7031 \\
		6 & 7550 & 7557 & 2775 & 3192.53 & 3478 & 2832 & 18.57 & 628.30 & 3301 & 2831 & 14.24 & 7078 \\
		7 & 7600 & 7734 & 2729 & 3223.23 & 3425 & 2831 & 17.34 & 642.02 & 3277 & 2830 & 13.64 & 7125 \\
		8 & 7650 & 7696 & 2797 & 3269.30 & 3490 & 2855 & 18.19 & 658.27 & 3336 & 2851 & 14.54 & 7171 \\
		9 & 7700 & 7716 & 2803 & 3325.65 & 3521 & 2862 & 18.72 & 703.98 & 3334 & 2860 & 14.22 & 7218 \\
		10 & 7750 & 7806 & 2807 & 3370.84 & 3534 & 2886 & 18.34 & 684.13 & 3344 & 2882 & 13.82 & 7265 \\
		11 & 7800 & 7804 & 2879 & 3428.42 & 3571 & 2932 & 17.89 & 704.58 & 3397 & 2932 & 13.69 & 7312 \\
		12 & 7850 & 7884 & 2851 & 3499.23 & 3556 & 2913 & 18.08 & 705.05 & 3363 & 2910 & 13.47 & 7359 \\
		13 & 7900 & 7932 & 2889 & 3584.14 & 3588 & 2947 & 17.87 & 727.14 & 3416 & 2946 & 13.76 & 7406 \\
		14 & 7950 & 7982 & 2949 & 6376.02 & 3641 & 3004 & 17.50 & 753.24 & 3481 & 3004 & 13.70 & 7453 \\
		15 & 8000 & 8126 & 2882 & 6828.32 & 3624 & 2966 & 18.16 & 758.84 & 3469 & 2964 & 14.56 & 7500 \\
		16 & 9150 & 10443603 & 3 & 6762.49 & 21 & 20 & 4.76 & 17.71 & 22 & 19 & 13.64 & 6485 \\
		17 & 9200 & 10558172 & 3 & 6887.40 & 21 & 20 & 4.76 & 18.11 & 22 & 19 & 13.64 & 6489 \\
		18 & 9250 & 10673366 & 3 & 7020.20 & 20 & 19 & 5.00 & 17.71 & 21 & 19 & 9.52 & 6523 \\
		19 & 9300 & 14357729 & 3 & 8222.71 & 20 & 18 & 10.00 & 21.33 & 21 & 17 & 19.05 & 5703 \\
		20 & 9350 & 14540834 & 2 & 8380.73 & 16 & 15 & 6.25 & 18.87 & 18 & 15 & 16.67 & 5700 \\
		21 & 9400 & 11001589 & 3 & 7428.46 & 23 & 21 & 8.70 & 21.87 & 25 & 21 & 16.00 & 6579 \\
		22 & 9450 & 14825554 & 3 & 8755.81 & 19 & 17 & 10.53 & 20.95 & 20 & 16 & 20.00 & 5742 \\
		23 & 9500 & 14983162 & 3 & 8857.65 & 20 & 18 & 10.00 & 25.09 & 20 & 16 & 20.00 & 5791 \\
		24 & 9550 & 15141604 & 3 & 8974.30 & 20 & 18 & 10.00 & 26.14 & 24 & 18 & 25.00 & 5773 \\
		25 & 9600 & 15329626 & 2 & 9114.64 & 16 & 15 & 6.25 & 19.76 & 18 & 16 & 11.11 & 5800 \\
		26 & 9650 & 15460987 & 3 & 10186.50 & 20 & 18 & 10.00 & 22.17 & 20 & 17 & 15.00 & 5790 \\
		27 & 9700 & 15621929 & 3 & 10488.80 & 19 & 17 & 10.53 & 22.48 & 20 & 16 & 20.00 & 5832 \\
		28 & 9750 & 15812901 & 2 & 10639.20 & 16 & 15 & 6.25 & 19.54 & 17 & 15 & 11.76 & 5842 \\
		29 & 9800 & 11959739 & 3 & 12332.80 & 23 & 21 & 8.70 & 26.78 & 28 & 21 & 25.00 & 6805 \\
		30 & 12600 & 39630183 & 2 & 30335.40 & 12 & 11 & 8.33 & 37.34 & 12 & 11 & 8.33 & 5406 \\
		31 & 12650 & 39945571 & 2 & 30143.80 & 13 & 12 & 7.69 & 44.06 & 14 & 11 & 21.43 & 5419 \\
		32 & 12700 & 40262208 & 2 & 35540.20 & 13 & 12 & 7.69 & 43.48 & 14 & 11 & 21.43 & 5436 \\
		33 & 12750 & 40580096 & 2 & 40648.00 & 13 & 12 & 7.69 & 44.04 & 14 & 11 & 21.43 & 5437 \\
		34 & 12800 & 40841707 & 3 & 36502.20 & 15 & 13 & 13.33 & 53.28 & 16 & 13 & 18.75 & 5438 \\
		35 & 12850 & 41219621 & 2 & 37844.20 & 13 & 12 & 7.69 & 41.89 & 13 & 12 & 7.69 & 5450 \\
		36 & 12900 & 41541258 & 2 & 39178.50 & 12 & 11 & 8.33 & 38.03 & 12 & 11 & 8.33 & 5448 \\
		37 & 12950 & 41864146 & 2 & 41767.80 & 12 & 11 & 8.33 & 37.53 & 12 & 11 & 8.33 & 5448 \\
		38 & 13000 & 42188283 & 2 & 43198.60 & 12 & 11 & 8.33 & 41.69 & 13 & 10 & 23.08 & 5433 \\
		39 & 13100 & 42840308 & 2 & 42710.00 & 12 & 11 & 8.33 & 42.70 & 13 & 10 & 23.08 & 5446 \\
		40 & 13150 & 43168196 & 2 & 40698.00 & 12 & 11 & 8.33 & 41.15 & 12 & 11 & 8.33 & 5442 \\ \hline
		\caption{Results of Algorithm H1 and H2 for the randomly generated graphs.}
		\label{table1}
	\end{longtable}
\end{scriptsize}

In Table \ref{table2} we compare the quality of the solutions obtained
by Algorithms H1 and H2 before the purification stage, represented by the parameters 
$|D_{H1}|$ and $|D_{H2}|$, respectively.  

\begin{footnotesize}
\begin{longtable}[c]{ccc}
	
	\cline{1-3}
	& \multicolumn{1}{c}{\textbf{Instances}} & \multicolumn{1}{c}{\textbf{\%}} \\ \hline
	\endhead
	\multicolumn{1}{l}{\textbf{$|D_{H1}|=|D_{H2}|$}} & 781 & 34.19 \\ 
	\multicolumn{1}{l}{\textbf{$|D_{H1}|>|D_{H2}|$}} & 365 & 15.98 \\ 
	\multicolumn{1}{l}{\textbf{$|D_{H1}|<|D_{H2}|$}} & 1138 & 49.83 \\ \hline
	
	\caption{Comparative analysis of heuristics H1 and H2 before the purification}
	\label{table2}
\end{longtable}
\end{footnotesize}

The tables below present comparative analysis of the heuristics after the purification. 
As we can see, in 100\% of the instances, the solutions delivered by Algorithm H2 
were purified. Algorithm H1 delivered minimal global dominating sets for 7.48\% of 
the instances (note that a minimal dominating set cannot be purified). 
Both algorithms halted within a few seconds for the instances with up to 1000 vertices.
Algorithm H2 turned out to be faster, delivering solutions within 
a few seconds for most of the instances, while for the largest instance with 
14 100 vertices, it hated within 50 seconds. Algorithm H1 took about 
11 hours for this instance. In average, among the instances that were solved by the both 
algorithms,  Algorithm H2 was about 88 times faster. 

\begin{footnotesize}
\begin{longtable}[c]{lcc}
	\cline{1-3}
	& \multicolumn{1}{l}{\textbf{Instances}} & \multicolumn{1}{l}{\textbf{\%}} \\ \hline
	\endhead
	\multicolumn{1}{l}{\textbf{$|D_{H1}^*|=|D_{H2}^*|$}} & 1286 & 56.31 \\ 
	\multicolumn{1}{l}{\textbf{$|D_{H1}^*|>|D_{H2}^*|$}} & 905 & 39.62 \\
	\multicolumn{1}{l}{\textbf{$|D_{H1}^*|<|D_{H2}^*|$}} & 93 & 4.07 \\ \hline
	\caption{Comparative analysis of heuristics H1 and H2 after the	purification}
	\label{table3}
\end{longtable}
\end{footnotesize}

The observed drastic difference in the execution times of Algorithms H1 and H2  
can be reduced by simply omitting the initial iteration $h=0$ of Algorithms H2, 
the most time consuming part of that algorithm. The execution time 
of the modified Algorithm H1 in average, has reduced almost 160 time. As to the
quality of the solutions, the original version gave better solutions in 18.43\% of 
the tested instances. Table \ref{table4} (Table \ref{table5}, respectively) gives more detailed comparisons before (after, respectively) application of the purification procedure; 
$|D_{H1}|$, $|D_{H1_{mod}}|$, $|D^*_{H1}|$, $|D^*_{H1_{mod}}|$ stand for the cardinality
of the corresponding global dominating sets, before and after the purification.

\begin{footnotesize}
\begin{longtable}[c]{ccc}
	\cline{1-3}
	& \multicolumn{1}{c}{\textbf{Instances}} & \multicolumn{1}{c}{\textbf{\%}} \\ \hline
	\endhead
	\multicolumn{1}{l}{\textbf{$|D_{H1}|=|D_{H1_{mod}}|$}} & 1558 & 68.21 \\ 
	\multicolumn{1}{l}{\textbf{$|D_{H1}|>|D_{H1_{mod}}|$}} & 190 & 8.32 \\ 
	\multicolumn{1}{l}{\textbf{$|D_{H1}|<|D_{H1_{mod}}|$}} & 536 & 23.47 \\ \hline
	
	\caption{Comparative analysis of the two versions of heuristic H1 before the
		purification}
	\label{table4}
\end{longtable} 
\end{footnotesize}

\begin{footnotesize}
\begin{longtable}[c]{lcc}
	\cline{1-3}
	& \multicolumn{1}{l}{\textbf{Instances}} & \multicolumn{1}{l}{\textbf{\%}} \\ \hline
	\endhead
	\multicolumn{1}{l}{\textbf{$|D_{H1}^*|=|D_{H1_{mod}}^*|$}} & 1660 & 72.68 \\ 
	\multicolumn{1}{l}{\textbf{$|D_{H1}^*|>|D_{H1_{mod}}^*|$}} &  203 & 8.89 \\
	\multicolumn{1}{l}{\textbf{$|D_{H1}^*|<|D_{H1_{mod}}^*|$}} &  421 & 18.43 \\ \hline
	\caption{Comparative analysis of the two versions of heuristic H1 after the
		purification}
	\label{table5}
\end{longtable}
\end{footnotesize}

Recall that Algorithms H3 uses Algorithm H1 as a subroutine. 
Above we considered a modification  of the latter algorithm, 
which resulted in a much faster performance but a bit worst
solution quality.  We carried out similar modifications in Algorithm H3, 
and we got similar outcome.  Table \ref{table6} gives related results for 
about 40 selected instances. Among the instances that were solved by 
Algorithm H3 and its modification, in average, the modified heuristic  was about 83 times faster. This drastic difference was achieved by just eliminating the initial 
iteration of Algorithm H3 (H1). Compared to 
Algorithm H2, in average, modified Algorithm H3 was about $ 5$ times slower. 
The global dominating sets delivered by Algorithm H3 and its modification were
purified in 93.15\% and 93.22\% of the instances, respectively (with an average
of approximately 30 reduced vertices). Tables \ref{table8} and \ref{table9} give
a comparative analysis of the two versions of the heuristic before and after 
the purification stage, respectively.

\begin{tiny}
	\begin{longtable}[c]{ccccccccccccc}
		\hline
		\multirow{2}{*}{No.} & \multirow{2}{*}{$|V|$} & \multirow{2}{*}{$|E|$} & \multirow{2}{*}{L} & \multicolumn{4}{c}{Algorithm H3} & \multicolumn{4}{c}{Modified Algorithm H3} & \multirow{2}{*}{U} \\ \cline{5-12}
		&  &  &  & Time(s) & $|D_{H3}|$ & $|D_{H3}^*|$ & \% Purified & Time(s) & $|D_{H3_{mod}}|$ & $|D_{H3_{mod}}^*|$ & \% Purified &  \\ \hline
		\endhead
		\hline
		\endfoot
		\endlastfoot
		1 & 7300 & 7311 & 2663 & 3934.11 & 3158 & 2706 & 14.31 & 1118.17 & 3158 & 2706 & 14.31 & 6843 \\
		2 & 7350 & 7474 & 2621 & 4262.89 & 3174 & 2722 & 14.24 & 1114.77 & 3174 & 2722 & 14.24 & 6890 \\
		3 & 7400 & 7474 & 2677 & 4306.95 & 3200 & 2740 & 14.38 & 1150.65 & 3200 & 2740 & 14.38 & 6937 \\
		4 & 7450 & 7497 & 2695 & 4731 & 3217 & 2755 & 14.36 & 1170.54 & 3217 & 2755 & 14.36 & 6984 \\
		5 & 7500 & 7535 & 2762 & 4862.41 & 3270 & 2822 & 13.70 & 1209.28 & 3270 & 2822 & 13.70 & 7031 \\
		6 & 7550 & 7557 & 2775 & 5244.28 & 3298 & 2831 & 14.16 & 1242.61 & 3298 & 2831 & 14.16 & 7078 \\
		7 & 7600 & 7734 & 2729 & 4510.99 & 3271 & 2829 & 13.51 & 1261.24 & 3271 & 2829 & 13.51 & 7125 \\
		8 & 7650 & 7696 & 2797 & 3370.89 & 3331 & 2851 & 14.41 & 1287.06 & 3331 & 2851 & 14.41 & 7171 \\
		9 & 7700 & 7716 & 2803 & 3418.88 & 3331 & 2860 & 14.14 & 1301.44 & 3331 & 2860 & 14.14 & 7218 \\
		10 & 7750 & 7806 & 2807 & 3544.64 & 3341 & 2882 & 13.74 & 1328.39 & 3342 & 2882 & 13.76 & 7265 \\
		11 & 7800 & 7804 & 2879 & 3586.43 & 3395 & 2932 & 13.64 & 1359.95 & 3395 & 2932 & 13.64 & 7312 \\
		12 & 7850 & 7884 & 2851 & 3645.01 & 3359 & 2910 & 13.37 & 1395.35 & 3359 & 2910 & 13.37 & 7359 \\
		13 & 7900 & 7932 & 2889 & 3708.27 & 3415 & 2946 & 13.73 & 1436.03 & 3415 & 2946 & 13.73 & 7406 \\
		14 & 7950 & 7982 & 2949 & 6423.91 & 3479 & 3004 & 13.65 & 1710.06 & 3479 & 3004 & 13.65 & 7453 \\
		15 & 8000 & 8126 & 2882 & 6742.18 & 3466 & 2964 & 14.48 & 1762.53 & 3466 & 2964 & 14.48 & 7500 \\
		16 & 9150 & 10443603 & 3 & 8849.74 & 21 & 20 & 4.76 & 24.7761 & 21 & 20 & 4.76 & 6485 \\
		17 & 9200 & 10558172 & 3 & 8219.12 & 21 & 20 & 4.76 & 24.461 & 21 & 20 & 4.76 & 6489 \\
		18 & 9250 & 10673366 & 3 & 8367.25 & 20 & 19 & 5.00 & 23.889 & 20 & 19 & 5.00 & 6523 \\
		19 & 9300 & 14357729 & 3 & 9683.22 & 20 & 18 & 10.00 & 27.0218 & 19 & 17 & 10.53 & 5703 \\
		20 & 9350 & 14540834 & 2 & 9899.63 & 16 & 15 & 6.25 & 24.9821 & 17 & 16 & 5.88 & 5700 \\
		21 & 9400 & 11001589 & 3 & 8808.05 & 23 & 21 & 8.70 & 27.4256 & 24 & 22 & 8.33 & 6579 \\
		22 & 9450 & 14825554 & 3 & 10319.8 & 19 & 17 & 10.53 & 27.8673 & 19 & 17 & 10.53 & 5742 \\
		23 & 9500 & 14983162 & 3 & 10548.7 & 20 & 18 & 10.00 & 27.7427 & 19 & 17 & 10.53 & 5791 \\
		24 & 9550 & 15141604 & 3 & 10751.4 & 20 & 18 & 10.00 & 29.1597 & 20 & 18 & 10.00 & 5773 \\
		25 & 9600 & 15329626 & 2 & 10890.4 & 16 & 15 & 6.25 & 28.2489 & 17 & 16 & 5.88 & 5800 \\
		26 & 9650 & 15460987 & 3 & 10941.8 & 20 & 18 & 10.00 & 29.9306 & 20 & 18 & 10.00 & 5790 \\
		27 & 9700 & 15621929 & 3 & 15073.5 & 19 & 17 & 10.53 & 31.4752 & 19 & 17 & 10.53 & 5832 \\
		28 & 9750 & 15812901 & 2 & 15339.3 & 16 & 15 & 6.25 & 28.8936 & 16 & 15 & 6.25 & 5842 \\
		29 & 9800 & 11959739 & 3 & 13978.7 & 23 & 21 & 8.70 & 31.5612 & 23 & 21 & 8.70 & 6805 \\
		30 & 12600 & 39630183 & 2 & 25428.1 & 12 & 11 & 8.33 & 38.1663 & 13 & 11 & 15.38 & 5406 \\
		31 & 12650 & 39945571 & 2 & 25952.5 & 13 & 12 & 7.69 & 38.9686 & 13 & 12 & 7.69 & 5419 \\
		32 & 12700 & 40262208 & 2 & 26387.9 & 13 & 12 & 7.69 & 36.3431 & 12 & 11 & 8.33 & 5436 \\
		33 & 12750 & 40580096 & 2 & 26763.6 & 13 & 12 & 7.69 & 44.107 & 12 & 11 & 8.33 & 5437 \\
		34 & 12800 & 40841707 & 3 & 31395 & 15 & 13 & 13.33 & 47.6085 & 16 & 14 & 12.50 & 5438 \\
		35 & 12850 & 41219621 & 2 & 31919.1 & 13 & 12 & 7.69 & 39.0515 & 12 & 11 & 8.33 & 5450 \\
		36 & 12900 & 41541258 & 2 & 32506.9 & 12 & 11 & 8.33 & 44.6907 & 13 & 12 & 7.69 & 5448 \\
		37 & 12950 & 41864146 & 2 & 34730.8 & 12 & 11 & 8.33 & 41.749 & 13 & 12 & 7.69 & 5448 \\
		38 & 13000 & 42188283 & 2 & 33157.2 & 12 & 11 & 8.33 & 39.7126 & 12 & 11 & 8.33 & 5433 \\
		39 & 13050 & 42455020 & 3 & 33496.8 & 15 & 13 & 13.33 & 47.6886 & 15 & 13 & 13.33 & 5450 \\
		40 & 13100 & 42840308 & 2 & 33583.9 & 12 & 11 & 8.33 & 39.2545 & 12 & 11 & 8.33 & 5446 \\
		41 & 13150 & 43168196 & 2 & 34047 & 12 & 11 & 8.33 & 39.0421 & 12 & 11 & 8.33 & 5442 \\ \hline
		\caption{Results for the two versions of heuristic H3 for 41
			randomly generated instances}
		\label{table6}
	\end{longtable}
\end{tiny}


\begin{footnotesize}
\begin{longtable}[c]{ccc}
	\cline{1-3}
	& \multicolumn{1}{c}{\textbf{Instances}} & \multicolumn{1}{c}{\textbf{\%}} \\ \hline
	\endhead
	\multicolumn{1}{l}{\textbf{$|D_{H3}|=|D_{H3_{mod}}|$}} & 1630 & 71.36 \\ 
	\multicolumn{1}{l}{\textbf{$|D_{H3}|>|D_{H3_{mod}}|$}} &  179 &  7.84 \\ 
	\multicolumn{1}{l}{\textbf{$|D_{H3}|<|D_{H3_{mod}}|$}} &  475 & 20.80 \\ \hline
	
	\caption{Comparative analysis of the two versions of heuristic H3 before the
	purification}
	\label{table8}
\end{longtable}
\end{footnotesize}

\begin{footnotesize}
\begin{longtable}[c]{lcc}
	\cline{1-3}
	& \multicolumn{1}{l}{\textbf{Instances}} & \multicolumn{1}{l}{\textbf{\%}} \\ \hline
	\endhead
	\multicolumn{1}{l}{\textbf{$|D_{H3}^*|=|D_{H3_{mod}}^*|$}} & 1686 & 73.82 \\ 
	\multicolumn{1}{l}{\textbf{$|D_{H3}^*|>|D_{H3_{mod}}^*|$}} &  189 &  8.27 \\
	\multicolumn{1}{l}{\textbf{$|D_{H3}^*|<|D_{H3_{mod}}^*|$}} &  409 & 17.91 \\ \hline
	\caption{Comparative analysis of the two versions of heuristic H3 after the
	purification}
	\label{table9}
\end{longtable}
\end{footnotesize}

{\bf Summary.} We summarize the obtained results for our heuristics. 
Before the application of the purification procedure,
solutions with equal cardinality were obtained in 
31\% of the analyzed instances. For the remaining instances,
Algorithms H1, H2, and H3 obtained the best solutions
in 70.54\%, 8.69\%, and 20.77\% of the tested instances, respectively.
When applying the purification procedure to the solutions 
delivered by Algorithm H2, the corresponding purified solutions
were better than ones constructed by Algorithms H1 and H3.  
In average, the purification procedure reduced the size of the solutions delivered 
by Algorithms H1, H2, and H3 by 16.03\%, 21.58\% and 15.34\%, respectively, see
Table \ref{table7} for a detailed comparative analysis. 

\begin{footnotesize}
\begin{longtable}[c]{lcc}
	
	\cline{1-3}
	& \multicolumn{1}{l}{\textbf{Instances}} & \multicolumn{1}{l}{\textbf{\%}} \\ \hline
	\endhead
	\multicolumn{1}{l}{\textbf{All}} & 1233 & 53.98 \\ 
	\multicolumn{1}{l}{\textbf{H1}} &   121 &  5.30 \\
	\multicolumn{1}{l}{\textbf{H2}} &   887 & 38.84 \\
	\multicolumn{1}{l}{\textbf{H3}} &    43 &  1.88 \\ \hline
	\caption{Comparative analysis of heuristics H1,H2 and H3 after the 	purification}
	\label{table7}
\end{longtable}
\end{footnotesize}

Comparing the size of the best obtained dominant  sets with 
the upper bound $ U $, $\min \{|D_{H1}^*|,|D_{H2}^*|,|D_{H3}^*|\}$ represents about $ 10\% $ 
of $ U $ in average, i.e., an improvement of 90\% over this upper bound. The lower bound 
$L$ represents, in average, $ 1/3 $ of the cardinality of the obtained solutions. 
The complete experimental  data can be found at \cite{Parra2022}.

\section{Conclusions}

We proposed the first exact and approximation algorithms for the classical global
domination problem. The exact algorithms were able to solve the existing benchmark instances
with up to 1098 vertices, while the heuristics found high-quality solutions for
the largest existing benchmark instances with up to 14 000 vertices. We  
showed that the global domination problem remains $NP$-hard for quite particular  
families of graphs and gave some families of graphs where our heuristics find an 
optimal solution. As to the future work, we believe that our line of research
can be extended for more complex domination problems.


\section*{Conflict of interest}
The authors declare there is no conflict of interest.

%


\bibliography{mybibfile_global}

\begin{thebibliography}{19}
\expandafter\ifx\csname natexlab\endcsname\relax\def\natexlab#1{#1}\fi
\providecommand{\url}[1]{\texttt{#1}}
\providecommand{\href}[2]{#2}
\providecommand{\path}[1]{#1}
\providecommand{\DOIprefix}{doi:}
\providecommand{\ArXivprefix}{arXiv:}
\providecommand{\URLprefix}{URL: }
\providecommand{\Pubmedprefix}{pmid:}
\providecommand{\doi}[1]{\href{http://dx.doi.org/#1}{\path{#1}}}
\providecommand{\Pubmed}[1]{\href{pmid:#1}{\path{#1}}}
\providecommand{\bibinfo}[2]{#2}
\ifx\xfnm\relax \def\xfnm[#1]{\unskip,\space#1}\fi
\bibitem[{Abu-Khzam \& Lamaa(2018)}]{abu2018}
\bibinfo{author}{Abu-Khzam, F.~N.}, \& \bibinfo{author}{Lamaa, K.}
  (\bibinfo{year}{2018}).
\newblock \bibinfo{title}{Efficient heuristic algorithms for positive-influence
  dominating set in social networks}.
\newblock In {\it \bibinfo{booktitle}{IEEE INFOCOM 2018-IEEE Conference on
  Computer Communications Workshops (INFOCOM WKSHPS)}\/} (pp.
  \bibinfo{pages}{610--615}).
\newblock \bibinfo{organization}{IEEE}.
\newblock \DOIprefix\doi{https://doi.org/10.1109/INFCOMW.2018.8406851}.
\bibitem[{Bertossi(1984)}]{Bertossi}
\bibinfo{author}{Bertossi, A.~A.} (\bibinfo{year}{1984}).
\newblock \bibinfo{title}{Dominating sets for split and bipartite graphs}.
\newblock {\it \bibinfo{journal}{Information processing letters}\/},  {\it
  \bibinfo{volume}{19}\/}, \bibinfo{pages}{37--40}.
  \DOIprefix\doi{https://doi.org/10.1016/0020-0190(84)90126-1}.
\bibitem[{Brigham \& Dutton(1990)}]{Dutton}
\bibinfo{author}{Brigham, R.~C.}, \& \bibinfo{author}{Dutton, R.~D.}
  (\bibinfo{year}{1990}).
\newblock \bibinfo{title}{Factor domination in graphs}.
\newblock {\it \bibinfo{journal}{Discrete Mathematics}\/},  {\it
  \bibinfo{volume}{86}\/}, \bibinfo{pages}{127--136}.
  \DOIprefix\doi{https://doi.org/10.1016/0012-365X(90)90355-L}.
\bibitem[{Colombi et~al.(2017)Colombi, Mansini \& Savelsbergh}]{COLOMBI}
\bibinfo{author}{Colombi, M.}, \bibinfo{author}{Mansini, R.}, \&
  \bibinfo{author}{Savelsbergh, M.} (\bibinfo{year}{2017}).
\newblock \bibinfo{title}{The generalized independent set problem: Polyhedral
  analysis and solution approaches}.
\newblock {\it \bibinfo{journal}{European Journal of Operational Research}\/},
  {\it \bibinfo{volume}{260}\/}, \bibinfo{pages}{41--55}.
  \DOIprefix\doi{https://doi.org/10.1016/j.ejor.2016.11.050}.
\bibitem[{Desormeaux et~al.(2015)Desormeaux, Gibson \& Haynes}]{Haynes}
\bibinfo{author}{Desormeaux, W.~J.}, \bibinfo{author}{Gibson, P.~E.}, \&
  \bibinfo{author}{Haynes, T.~W.} (\bibinfo{year}{2015}).
\newblock \bibinfo{title}{Bounds on the global domination number}.
\newblock {\it \bibinfo{journal}{Quaestiones Mathematicae}\/},  {\it
  \bibinfo{volume}{38}\/}, \bibinfo{pages}{563--572}.
  \DOIprefix\doi{https://doi.org/10.2989/16073606.2014.981728}.
\bibitem[{Doreian \& Conti(2012)}]{social2012}
\bibinfo{author}{Doreian, P.}, \& \bibinfo{author}{Conti, N.}
  (\bibinfo{year}{2012}).
\newblock \bibinfo{title}{Social context, spatial structure and social network
  structure}.
\newblock {\it \bibinfo{journal}{Social networks}\/},  {\it
  \bibinfo{volume}{34}\/}, \bibinfo{pages}{32--46}.
  \DOIprefix\doi{https://doi.org/10.1016/j.socnet.2010.09.002}.
\bibitem[{Enciso \& Dutton(2008)}]{Enciso}
\bibinfo{author}{Enciso, R.~I.}, \& \bibinfo{author}{Dutton, R.~D.}
  (\bibinfo{year}{2008}).
\newblock \bibinfo{title}{Global domination in planar graphs}.
\newblock {\it \bibinfo{journal}{J. Combin. Math. Combin. Comput}\/},  {\it
  \bibinfo{volume}{66}\/}, \bibinfo{pages}{273--278}. \URLprefix
  \url{https://citeseerx.ist.psu.edu/viewdoc/summary?doi=10.1.1.535.9636}.
\bibitem[{Foldes \& Hammer(1977)}]{Foldes}
\bibinfo{author}{Foldes, S.}, \& \bibinfo{author}{Hammer, P.~L.}
  (\bibinfo{year}{1977}).
\newblock \bibinfo{title}{Split graphs having dilworth number two}.
\newblock {\it \bibinfo{journal}{Canadian Journal of Mathematics}\/},  {\it
  \bibinfo{volume}{29}\/}, \bibinfo{pages}{666--672}.
  \DOIprefix\doi{https://doi.org/10.4153/CJM-1977-069-1}.
\bibitem[{Garey \& Johnson(1979)}]{Garey}
\bibinfo{author}{Garey, M.~R.}, \& \bibinfo{author}{Johnson, D.~S.}
  (\bibinfo{year}{1979}).
\newblock {\it \bibinfo{title}{Computers and intractability}\/} volume
  \bibinfo{volume}{174}.
\newblock \bibinfo{publisher}{freeman San Francisco}.
\bibitem[{Haynes(2017)}]{haynes2017domination}
\bibinfo{author}{Haynes, T.} (\bibinfo{year}{2017}).
\newblock {\it \bibinfo{title}{Domination in Graphs: Volume 2: Advanced
  Topics}\/}.
\newblock \bibinfo{publisher}{Routledge}.
\bibitem[{Inza et~al.(2023)Inza, Vakhania, Almira \& Mira}]{ejor2023exact}
\bibinfo{author}{Inza, E.~P.}, \bibinfo{author}{Vakhania, N.},
  \bibinfo{author}{Almira, J. M.~S.}, \& \bibinfo{author}{Mira, F. A.~H.}
  (\bibinfo{year}{2023}).
\newblock \bibinfo{title}{Exact and heuristic algorithms for the domination
  problem}.
\newblock {\it \bibinfo{journal}{European Journal of Operational Research}\/},
  . \DOIprefix\doi{https://doi.org/10.1016/j.ejor.2023.08.033}.
\bibitem[{Jovanovic et~al.(2023)Jovanovic, Sanfilippo \& Voß}]{JOVANOVIC}
\bibinfo{author}{Jovanovic, R.}, \bibinfo{author}{Sanfilippo, A.~P.}, \&
  \bibinfo{author}{Voß, S.} (\bibinfo{year}{2023}).
\newblock \bibinfo{title}{Fixed set search applied to the clique partitioning
  problem}.
\newblock {\it \bibinfo{journal}{European Journal of Operational Research}\/},
  {\it \bibinfo{volume}{309}\/}, \bibinfo{pages}{65--81}.
  \DOIprefix\doi{https://doi.org/10.1016/j.ejor.2023.01.044}.
\bibitem[{Meek \& {Gary Parker}(1994)}]{MEEK}
\bibinfo{author}{Meek, D.}, \& \bibinfo{author}{{Gary Parker}, R.}
  (\bibinfo{year}{1994}).
\newblock \bibinfo{title}{A graph approximation heuristic for the vertex cover
  problem on planar graphs}.
\newblock {\it \bibinfo{journal}{European Journal of Operational Research}\/},
  {\it \bibinfo{volume}{72}\/}, \bibinfo{pages}{588--597}.
  \DOIprefix\doi{https://doi.org/10.1016/0377-2217(94)90425-1}.
\bibitem[{Merris(2003)}]{merris2003}
\bibinfo{author}{Merris, R.} (\bibinfo{year}{2003}).
\newblock \bibinfo{title}{Split graphs}.
\newblock {\it \bibinfo{journal}{European Journal of Combinatorics}\/},  {\it
  \bibinfo{volume}{24}\/}, \bibinfo{pages}{413--430}.
  \DOIprefix\doi{https://doi.org/10.1016/S0195-6698(03)00030-1}.
\bibitem[{Parra~Inza(2023)}]{Parra2022}
\bibinfo{author}{Parra~Inza, E.} (\bibinfo{year}{2023}).
\newblock \bibinfo{title}{Random graph (1)}.
\newblock {\it \bibinfo{journal}{Mendeley Data}\/},  {\it
  \bibinfo{volume}{V4}\/}.
  \DOIprefix\doi{https://doi.org/10.17632/rr5bkj6dw5.4}.
\bibitem[{Sampathkumar(1989)}]{Sampathkumar}
\bibinfo{author}{Sampathkumar, E.} (\bibinfo{year}{1989}).
\newblock \bibinfo{title}{The global domination number of a graph}.
\newblock {\it \bibinfo{journal}{J. Math. Phys. Sci}\/},  {\it
  \bibinfo{volume}{23}\/}, \bibinfo{pages}{377--385}.
\bibitem[{Tyshkevich \& Chernyak(1979)}]{Tyshkevich}
\bibinfo{author}{Tyshkevich, R.~I.}, \& \bibinfo{author}{Chernyak, A.~A.}
  (\bibinfo{year}{1979}).
\newblock \bibinfo{title}{Canonical partition of a graph defined by the degrees
  of its vertices}.
\newblock {\it \bibinfo{journal}{Isv. Akad. Nauk BSSR, Ser. Fiz.-Mat. Nauk}\/},
   {\it \bibinfo{volume}{5}\/}, \bibinfo{pages}{14--26}.
\bibitem[{Wang et~al.(2009)Wang, Camacho \& Xu}]{wang2009}
\bibinfo{author}{Wang, F.}, \bibinfo{author}{Camacho, E.}, \&
  \bibinfo{author}{Xu, K.} (\bibinfo{year}{2009}).
\newblock \bibinfo{title}{Positive influence dominating set in online social
  networks}.
\newblock In {\it \bibinfo{booktitle}{Combinatorial Optimization and
  Applications: Third International Conference, COCOA 2009, Huangshan, China,
  June 10-12, 2009. Proceedings 3}\/} (pp. \bibinfo{pages}{313--321}).
\newblock \bibinfo{organization}{Springer}.
\bibitem[{Wang et~al.(2011)Wang, Du, Camacho, Xu, Lee, Shi \& Shan}]{wang2011}
\bibinfo{author}{Wang, F.}, \bibinfo{author}{Du, H.}, \bibinfo{author}{Camacho,
  E.}, \bibinfo{author}{Xu, K.}, \bibinfo{author}{Lee, W.},
  \bibinfo{author}{Shi, Y.}, \& \bibinfo{author}{Shan, S.}
  (\bibinfo{year}{2011}).
\newblock \bibinfo{title}{On positive influence dominating sets in social
  networks}.
\newblock {\it \bibinfo{journal}{Theoretical Computer Science}\/},  {\it
  \bibinfo{volume}{412}\/}, \bibinfo{pages}{265--269}.
  \DOIprefix\doi{https://doi.org/10.1016/j.tcs.2009.10.001}.

\end{thebibliography}

\end{document}